\documentclass[pra,twocolumn,showpacs,preprintnumbers,amsmath,amssymb,superscriptaddress,longbibliography]{revtex4-2}

\usepackage{graphicx}
\usepackage{dcolumn}
\usepackage{bm}
\usepackage{color}
\usepackage{bbm}
\usepackage[colorlinks=true,citecolor=blue,linkcolor=blue,urlcolor=blue]{hyperref}
\usepackage{appendix}

\usepackage{xcolor}  
\usepackage{soul}       
\usepackage{cancel}

\begin{document}

\title{Estimating the performance boundary of Gottesman-Kitaev-Preskill codes and number-phase codes}

\author{Kai-Xuan Wen}
\affiliation{School of Physics, Sun Yat-sen University, Guangzhou 510275, China}
\author{Dong-Long Hu}
\email{2025511209@hust.edu.cn}
\affiliation{School of Physics and Institute for Quantum Science and Engineering,
Huazhong University of Science and Technology, Wuhan 430074, China}
\affiliation{School of Physics, Sun Yat-sen University, Guangzhou 510275, China}
\author{Shengyong Li}
\affiliation{Department of Automation, Tsinghua University, Beijing 100084, China}
\author{Ze-Liang Xiang}
\email{xiangzliang@mail.sysu.edu.cn}
\affiliation{School of Physics, Sun Yat-sen University, Guangzhou 510275, China}
\affiliation{State Key Laboratory of Optoelectronic Materials and Technologies, Sun Yat-sen University, Guangzhou 510275, China}

\date{\today}

\begin{abstract}
Bosonic quantum error-correcting codes encode logical information in a harmonic oscillator, with the Gottesman-Kitaev-Preskill (GKP) and number-phase (NP) codes representing two fundamentally different encoding paradigms. Although both have been extensively studied, it remains unclear under what physical noise conditions (including photon loss and dephasing) one encoding intrinsically outperforms the other. Here we estimate a quantitative performance boundary between GKP and NP codes under general photon loss-dephasing noise. By optimizing code parameters within each encoding family, we identify the noise regimes in which each code exhibits a fundamental advantage. In particular, we find that the crossover occurs when the dephasing strength is approximately two orders of magnitude smaller than the loss strength, revealing a sharp separation between operational regimes. Beyond this specific comparison, our work establishes a practical and extensible methodology for benchmarking bosonic codes and optimizing their parameters, providing concrete guidance for the experimental selection and deployment of bosonic encodings in realistic noise environments.
\end{abstract} 

\maketitle
% ---------------------------------------------------------------------------
% ---------------------------------------------------------------------------

\section{\label{sec:intro}INTRODUCTION}

Bosonic quantum error-correcting (QEC) codes~\cite{Braunstein1998QECforQV,gottesman2001encoding,Cochrane1999cat1,Chuang1997Bosoniccodes,Lloyd1998Analog,Hu2025NP,Grimsmo2020RotationCodes,Michael2016binomial,Leghtas2013cat2,Schlegel2022CatStates,Teoh2023Dualencoding,Hayden_2016,Niset2008Feasible,Bergmann2016NOONstates,Niu2018sym_ope,BRADY2024100496,Jain2024sphericalcodes,Bashmakova2025squeezedFock,Zeng2023ApproxRL,Zeng2025ApproxGKP} provide a powerful framework for protecting logical quantum information by exploiting the infinite-dimensional Hilbert space of a single harmonic oscillator. These hardware-efficient codes encode redundancy directly in continuous-variables (CVs) and have recently been experimentally demonstrated to exceed the break-even point of quantum error correction~\cite{Ofek2016catexp,Ni2023binomialexp,Sivak2023,Cai2024,Brock2025}.

In realistic CV platforms, decoherence of bosonic modes is dominated by photon loss and dephasing. Photon loss induces stochastic displacements in phase space, while dephasing corresponds to random phase rotations. Therefore, effective bosonic QEC requires encoding structures that are intrinsically robust against displacements, rotations, or both. 

Among existing proposals, the Gottesman-Kitaev-Preskill (GKP) code~\cite{gottesman2001encoding,Grimsmo2021GKPCode} and the number-phase (NP) code~\cite{Hu2025NP,Grimsmo2020RotationCodes} represent two paradigmatic bosonic lattice encodings distinguished by their symmetries. GKP codes encode logical information in a phase-space lattice protected by translational symmetry in position and momentum, rendering them particularly resilient to small displacement errors, which dominate in Gaussian noise channels arising from thermal noise and photon loss. In contrast, NP codes, including well-known cat codes~\cite{Cochrane1999cat1} and binomial codes~\cite{Michael2016binomial}, encode logical information on a lattice in the number-phase domain. Their structure endows them with intrinsic robustness against phase-rotation errors associated with dephasing, while simultaneously retaining nontrivial protection against photon loss through their number-space structure. 

These symmetry distinctions lead to qualitatively different noise-resilience characteristics and scaling behaviors. When noise channels are dominated by a single error type, increasing the mean photon number can monotonically suppress the corresponding correctable error~\cite{Albert2018PerformanceSingleMode}. Under general loss-dephasing noise, however, displacement and rotation errors compete, producing an intrinsic trade-off between different lattice codes. In this regime, both the code size and the lattice geometry determine performance. 

Despite extensive analytical insights into loss-dephasing channels~\cite{Ouyang2021NPtradeoff,Leviant2022quantumcapacity,Mele2024quantumcom_lossdephasing}, systematic quantitative comparisons of bosonic codes across broad noise regimes remain limited. Most previous works explore restricted parameter subsets, such as fixed lattice geometries or selected mean photon numbers, and evaluate performance under specific noise assumptions~\cite{Albert2018PerformanceSingleMode,totey2023performancerandombosonicrotation,Schlegel2022CatStates,Grimsmo2021GKPCode,Bashmakova2025squeezedFock}. As a result, the global performance landscape and the boundary separating regimes of intrinsic advantage have not been quantitatively established. 

In this work, we develop a scalable parameter-optimization framework that systematically maximizes the performance of bosonic QEC codes under general loss-dephasing noise. Treating the code performance as the optimization metric, our method employs an evolutionary algorithm to identify near-globally optimal parameter sets tailored to given noise conditions. Using this framework, we perform a large-scale, quantitative comparison between GKP and NP codes across a broad range of loss-dephasing strengths. Consistent with symmetry-based expectations, we find that GKP codes outperform NP codes in regimes dominated by photon loss, whereas NP codes exhibit superior resilience in dephasing-dominated regimes while maintaining nontrivial robustness against photon loss. Particularly, we numerically resolve their distinct performance regimes and estimate the boundary separating them. This performance boundary emerges when the dephasing is approximately two orders of magnitude smaller than the photon loss, delineating the crossover between displacement-protected and rotation-protected encoding. 

\begin{figure*}
    \centering
    \includegraphics[width=1\linewidth]{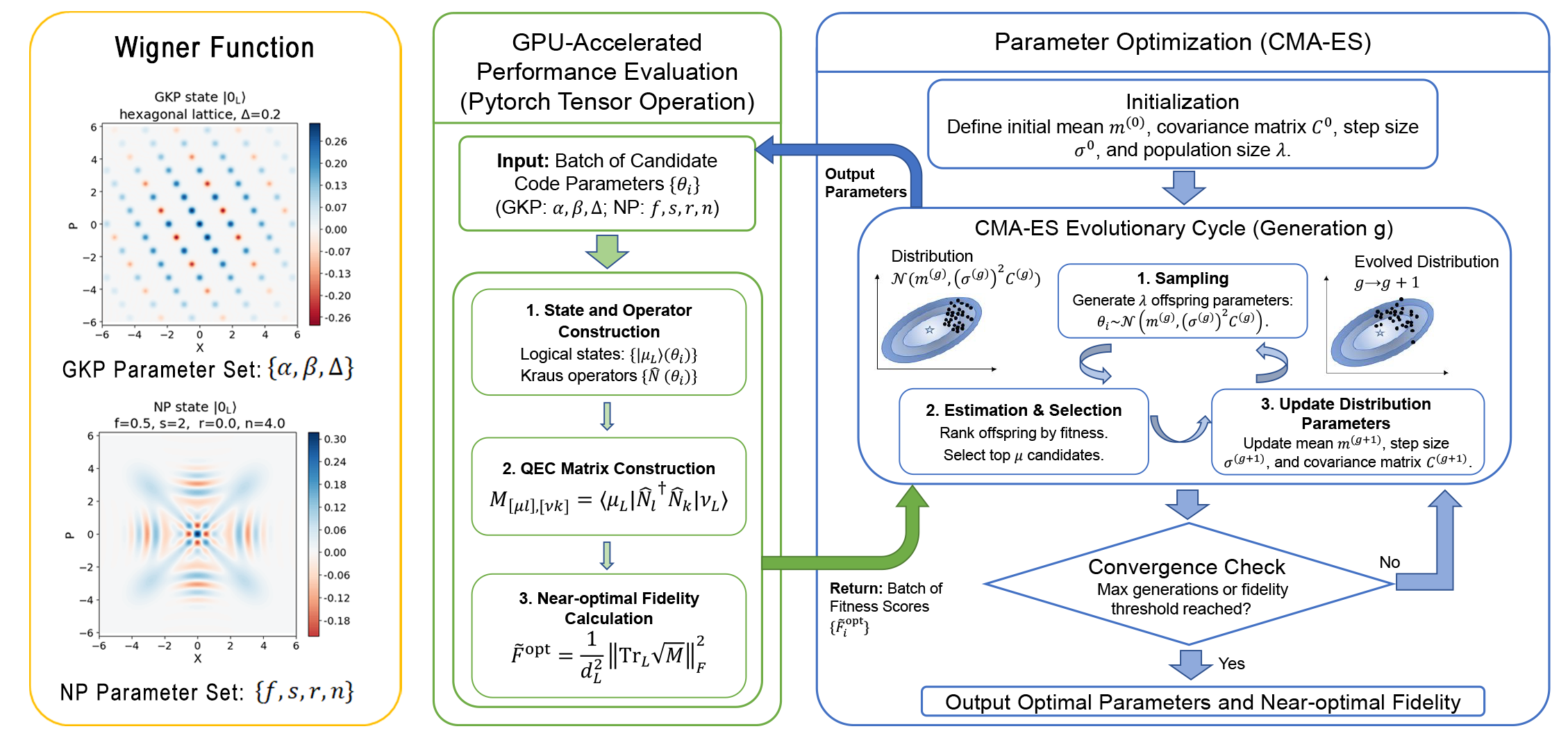}
    \caption{Overview of the bosonic lattice codes and the optimization workflow. The yellow panel shows the Wigner function of the logical state $|0_L\rangle$ for a hexagonal GKP code ($\Delta=0.2$) and the diamond NP code ($f=0.5,s=2,r=0,n=4$), together with the set of variational parameters. The green panel illustrates the GPU-accelerated evaluation of the near-optimal fidelity used as the performance metric. The blue panel summarizes the core mechanism of the covariance matrix adaptation evolution strategy (CMA-ES), showing how code parameters are iteratively updated to maximize performance. The green and blue rounded-corner arrows indicate the iterative feedback loop between fidelity evaluation and parameter optimization.}
    \label{fig:workflow}
\end{figure*}

The overall workflow of our optimization procedure is illustrated in Fig.~\ref{fig:workflow}. As a performance metric, we adopt the near-optimal fidelity~\cite{Zheng2024NearOptimal}, which inherits the two-sided bound of optimal fidelity and is a close approximation when the code performs well against the noise channel. Crucially, this metric captures full dependence of error-correction performance on both code parameters and noise strength, and can be evaluated directly from the QEC matrix. This structure allows us to leverage GPU-accelerated matrix operations, reducing the time required for a single fidelity evaluation to seconds and achieving one to two orders of magnitude speedup compared with conventional convex-optimization approaches~\cite{choi1975completely,kitaev1997quantum,schumacher1996entanglement,Reimpell2005IterativeOpt,reimpell2005optimal,Fletcher2007OptRecovery,watrous2009semidefinite,kosut2008robust,Audenaert2002OptSDP1,Mironowicz2024SDPandQI}. By combining GPU-accelerated fidelity evaluation with the covariance matrix adaptation evolution strategy (CMA-ES)~\cite{hansen2001completely,hansen2003reducing,Hansen2011CMAESES,hansen2014cmaes,2024CMAES} (for details see Sec.~\ref{sec:CMA-ES}), an evolutionary algorithm well suited for high-dimensional and nonconvex landscapes, our method efficiently explores the parameter space of bosonic codes under realistic noise models. Although practical constraints arise from Hilbert-space truncation and the increasing number of Kraus operators required at high energies and strong noise, the method nonetheless enables systematic exploration of a substantially broader performance landscape than previously accessible. 

% ---------------------------------------------------------------------------
% ---------------------------------------------------------------------------

This paper is organized as follows. In Sec.~\ref{sec:Framework}, we formulate the physical problem and detail the implementation of the method. Section~\ref{sec:results} presents the numerical results together with the corresponding physical insights and computational settings. Finally, Sec.~\ref{sec:summary} summarizes our main findings.
% ---------------------------------------------------------------------------
% ---------------------------------------------------------------------------

\section{Parameter optimization}
\label{sec:Framework}

In this section, we provide a detailed description of the simulation workflow illustrated in Fig.~\ref{fig:workflow}. We first introduce bosonic lattice codes and the near-optimal fidelity metric, then analyze computational complexity. We then formulate the optimization problem, describe the core principles of the evolutionary algorithm employed, and briefly discuss its robustness.

\subsection{Bosonic lattice codes}
\label{sec:bosonic codes}

Before presenting the specific simulation method, we first introduce the GKP and NP codes and specify the code parameters to be optimized.

\subsubsection{Gottesman-Kitaev-Preskill codes}
\label{sec:gkp_codes}

The GKP code encodes a logical qubit in the continuous degrees of freedom of a harmonic oscillator. Here, we focus on the simplest setting of a single logical qubit encoded in one bosonic mode. The ideal GKP states~\cite{gottesman2001encoding} are defined as simultaneous $+1$ eigenstates of the continuous-variable stabilizer group $\{\hat{S}^k_X, \hat{S}^l_Z\}$ for $k,l \in \mathbb{Z}$, with 
\begin{equation}
    \hat{S}_X =\bar{X}^2 = \hat{D}(2\alpha), \quad \hat{S}_Z = \bar{Z}^2 = \hat{D}(2\beta),
\end{equation}
and logical operators
\begin{equation}
    \bar{X}=\hat{D}(\alpha),\quad\bar{Z}=\hat{D}(\beta),
\end{equation}
where $\alpha,\beta \in \mathbb{C}$ satisfy 
\begin{equation}
    \beta\alpha^{*}-\beta^{*}\alpha = i\pi,
    \label{eq:GKP_condition}
\end{equation}
and the displacement operator is defined as $\hat{D}(\alpha)=e^{\alpha\hat{a}^\dagger-\alpha^{*}\hat{a}}$.

The logical state $|0_L\rangle \propto \sum^\infty _{k,l=-\infty} \hat{S}^k_X \bar{Z}^l |0\rangle$ is a simultaneous $+1$ eigenstate of the two stabilizer generators $\hat{S}_X, \hat{S}_Z$ and logical operator $\bar{Z}$. Equivalently, the logical codewords can be expressed as infinite superpositions of coherent states~\cite{Grimsmo2021GKPCode},
\begin{equation}
\begin{split}
|0_L\rangle &\propto \sum_{k,l=-\infty}^{\infty} e^{-i\pi kl}|2k\alpha + l\beta\rangle, \\
|1_L\rangle &\propto \sum_{k,l=-\infty}^{\infty} e^{-i\pi(kl+l/2)}|(2k+1)\alpha + l\beta\rangle, 
\end{split}
\label{eq:GKP}
\end{equation}
where the kets on the right-hand side denote coherent states.

Any pair $\alpha$, $\beta$ satisfying Eq.~\eqref{eq:GKP_condition} defines a valid lattice geometry of GKP codes. Among all choices, the hexagonal lattice maximizes symmetry and packing efficiency, corresponding to $$\alpha = \sqrt{\frac{\pi}{\sqrt{3}}}, \quad \beta = e^{2i\pi/3}\sqrt{\frac{\pi}{\sqrt{3}}}.$$ 
In our simulation, $\alpha$ and $\beta$ can be randomly chosen. 

The codewords in Eqs.~\eqref{eq:GKP} are ideal grid states with infinite energy and are therefore not physically realizable. In practice, one introduces a finite-energy approximation by applying a Gaussian envelope to the lattice,
\begin{equation}
    |\tilde{\mu}_L\rangle \propto e^{-\Delta^2 \hat{a}^\dagger \hat{a}}|\mu_L \rangle,
\end{equation}
where $\mu=0,1$ denote the two logical states and normalization is omitted for brevity. As $\Delta$ decreases, the peaks in phase space become sharper, leading to a broader momentum distribution and an increase of the mean photon number. In other words, the ideal limit corresponds to $\Delta \rightarrow 0$.

We therefore identify the set of GKP code parameters relevant to performance optimization as 
\begin{equation}
    \mathcal{T}_{\rm{GKP}}=\{\alpha,\beta,\Delta\}.
\end{equation}
In the following sections, we investigate how these parameters influence code performance under general loss-dephasing noise channels.

% ---------------------------------------------------------------------------

\subsubsection{Number-phase codes}

In analogy to the discrete translation symmetry underlying GKP codes in phase space, NP codes belong to a broader class of bosonic codes characterized by discrete translation symmetry in number-phase space. Typical members of this family include cat codes~\cite{Cochrane1999cat1}, binomial codes~\cite{Michael2016binomial} codes, and more general NP vortex codes with non-rectangle lattice geometries~\cite{Hu2025NP}. Here we briefly introduce the generalized number-phase code and its finite-energy approximation.

Whereas GKP codes employ displacement operators $\hat{D}(\alpha)=e^{\alpha\hat{a}^\dagger-\alpha^{*}\hat{a}}$ to generate translations in position-momentum space, translations in number-phase space are generated by the operators
\begin{equation}
    \hat{\mathcal{D}}(\vec{n}) = \exp(\frac{il\phi}{2})\hat{R}(\phi)\hat{\Sigma}_l,
\end{equation}
where $\hat{R}(\phi)$ is the phase-rotation operator with $\vec{n}=(l,\phi)$ for $l\in \mathbb{Z}$ and $\phi \in \mathbb{R}$, and $\hat{\Sigma}_l = \sum_n |n\rangle \langle n+l|$ is the number-translation operator. 

Generalized NP codes are defined as simultaneous $+1$ eigenstates of the stabilizer group $\{\hat{S}^k_X, \hat{S}^l_Z\}$ for $k,l \in \mathbb{Z}$, with 
\begin{equation}
    \hat{S}_X =\bar{X}^2 = \hat{\mathcal{D}}(2\vec{n}_x), \quad \hat{S}_Z = \bar{Z}^2 = \hat{\mathcal{D}}(2\vec{n}_z),
\end{equation}
where $\bar{X}=\hat{\mathcal{D}}(\vec{n}_x)$ and $\bar{Z}=\hat{\mathcal{D}}(\vec{n}_z)$ act as logical Pauli operators. The resulting codewords form a two-dimensional lattice in number-phase space, whose unit-cell area satisfies 
\begin{equation}\label{eq:NP_condition}
    |\vec{n}_x\times\vec{n}_z|=\pi.
\end{equation}

For NP codes, any pair ($\vec{n}_x$, $\vec{n}_z$) obeying Eq.~\eqref{eq:NP_condition} defines a valid lattice geometry. In this work, we adopt the parametrization
\begin{equation}
    \vec{n}_x=\left(s,\frac{f\pi}{s}\right),\quad \vec{n}_z=\left(0,\frac{\pi}{s}\right),
    \label{eq:NP_condition2}
\end{equation}
where $s$ is a positive integer setting the lattice spacing and $f\in[0,1)$ controls the lattice skewness. 

The idea NP codewords possess exact translational symmetry and are therefore non-normalized. To obtain physical states with finite energy, we introduce a normalized envelope $\{\theta_n\}$ satisfying $\sum_n |\theta_n|^2=1$. The approximate NP states can then be written in terms of sums of Fock states as 
\begin{equation}\label{eq:np_equation}
    |\pm\rangle_L(s, f, \theta_n) = \exp\left(-\frac{if\pi\hat{n}^2}{2s^2}\right) \sum_{n=0}^{\infty} (\pm)^n\theta_n|sn\rangle,
\end{equation}
with logical Z basis defined by $|\mu\rangle_{L} = (|+\rangle_{L} + (-)^{\mu}|-\rangle_{L})/\sqrt{2}$ with $\mu = 0,1$. The mean photon number of the approximate NP code state is
\begin{equation}
    \bar{n}_\text{code} = \sum_{n=0}^{\infty} |\theta_n|^2 sn < \infty,
\end{equation}
and the ideal NP limit is recovered when the infinite series $\{\theta_n\}$ becomes uniform.

While the lattice parameters $s$ and $f$ determine the geometric structure of the code in NP space, the envelope $\{\theta_n\}$ influences the resistance of the code to phase noise. In principle, infinitely many choices of $\{\theta_n\}$ are possible. Here, we restrict attention to envelopes derived from pure Gaussian states,
\begin{equation}
    \theta_n=\langle n|\alpha,r\rangle,\quad |\alpha,r\rangle=\hat{D}(\alpha)\hat{S}(r)|0\rangle,
\end{equation}
where $\hat{D}(\alpha)=e^{\alpha\hat{a}^\dagger-\alpha^{*}\hat{a}}$ and $\hat{S}(r)=\exp\left(\frac{r}{2}(\hat{a}^2-\hat{a}^{\dagger 2})\right)$ are the displacement and squeezing operators, respectively. For a displaced squeezed vacuum state, $\hat{D}(\alpha)\hat{S}(r)|0\rangle$, the parameters satisfy
\begin{equation}
    n = |\alpha|^2 + \sinh^2 r,
\end{equation}
where we emphasize that the mean photon number of the envelope state generally differs from that of the logical NP codewords. In the numerical analysis below, we therefore parametrize the envelope $\{\theta_n\}$ in terms of $n$ and $r$.

It is noticeable that the generalized NP codes defined in Eq.~\eqref{eq:np_equation} exhibit profound connections with mainstream bosonic QEC codes, including the well-known cat codes and binomial codes, which are special cases of rectangular-NP codes with different Fock amplitudes $\{\theta_n\}$.
% and the deciding distinctions are the different choices of $\{\theta_n\}$. 
For cat codes and binomial codes, their $\{\theta_n\}$ correspond to
\begin{equation}
    \theta_n^\textrm{cat}=\sqrt{\mathcal{N}}\langle sn|\alpha\rangle,\quad \theta_n^{\textrm{bin}}=\sqrt{2^{-N}{N \choose n}},
\end{equation}
with $\mathcal{N}$ a normalization constant.

To briefly summarize, the performance of an approximate NP code is governed by the parameter set
\begin{equation}
    \mathcal{T}_{\rm{NP}} = \{f,s,r,n \}.
\end{equation}
The enlarged parameter space provides substantial tunability beyond conventional cat and binomial codes, naturally framing NP encoding as a fidelity optimization problem over realistic noise channels.
% ---------------------------------------------------------------------------
% ---------------------------------------------------------------------------

\subsection{Error model}

To investigate the performance of QEC codes, we consider a generalized noise channel incorporating both photon loss and dephasing. $\mathcal{N}(\hat{\rho})$ is defined as the solution to the Lindblad master equation
\begin{equation}
    \frac{\partial \hat{\rho}(t)}{\partial t} = \gamma \mathcal{L}[\hat{a}]\hat{\rho}(t) + \kappa \mathcal{L}[\hat{n}]\hat{\rho}(t),
\end{equation}
evolved for a total time $t$. Here the Lindblad superoperator is $\mathcal{L}[\hat{A}]\hat{\rho}(t)=\hat{A}\hat{\rho}(t)\hat{A}^{\dagger}-1/2 \rho(t)\hat{A}^{\dagger}\hat{A}-1/2\hat{A}^{\dagger}\hat{A}\rho(t)$. The first term describes photon loss at rate $\gamma$, while the second term accounts for pure dephasing at rate $\kappa$.
In the results presented below, we adopt the dimensionless parameters $\gamma t$ and $\kappa t$ to represent the strengths of the simultaneous loss and dephasing rates.

The Kraus operator decomposition of $\mathcal{N}(\hat{\rho})$ is provided in Appendix~\ref{sec:kraus_ope}. Denoting the photon-loss Kraus operators $\{\hat{A}_k\}$ and the dephasing Kraus operators $\{\hat{B}_l\}$, the channel can be written as
\begin{equation}
    \mathcal{N}(\hat{\rho}) = \sum_{k,l\geq0}\hat{B}_l \hat{A}_k \hat{\rho} \hat{A}_k^\dagger \hat{B}_l^\dagger.
    \label{eq:N_form2}
\end{equation}
With the composite Kraus operators $\{\hat{N}_{k,l}\} = \{\hat{B}_l \hat{A}_k\}$, the QEC matrix for a code with $d_L$ logical codewords $\{|\mu_L\rangle\}$ is defined as
\begin{equation}
    M_{[\mu,(k,l)],[\nu,(k',l')]} =
    \langle \mu_L |
    \hat{N}_{k,l}^\dagger \hat{N}_{k',l'}
    |
    \nu_L \rangle.
\end{equation}
which compactly encodes how physical errors act within the logical subspace.

% ---------------------------------------------------------------------------

\subsection{Validity and efficiency analysis of near-optimal fidelity}

Evaluating the performance of QEC codes generally requires computing the channel fidelity under optimal recovery. This optimal channel fidelity $F^{\mathrm{opt}}$ can be obtained via a semidefinite program on a Choi matrix~\cite{choi1975completely,schumacher1996entanglement,Reimpell2005IterativeOpt,reimpell2005optimal,Fletcher2007OptRecovery}. As shown in Refs.~\cite{9317892,Zheng2024NearOptimal}, the computational cost scales as
\begin{equation}
\tilde{\mathcal{O}}[(d_L N)^{5.246}],
\end{equation} 
where $d_L$ denotes the logical code dimension, $N$ the truncated Fock-space dimension, and $\tilde{\mathcal{O}}[g(n)]\equiv\mathcal{O}[g(n) log^k g(n)]$. This scaling rapidly becomes prohibitive as either the code energy or the truncation dimension increases.

In contrast, the near-optimal channel fidelity $\tilde{F}^{\mathrm{opt}}$~\cite{Zheng2024NearOptimal} can be evaluated directly from its QEC matrix, reducing the computational cost to
\begin{equation}
\mathcal{O}[(d_L N_K)^3],
\end{equation}
where $N_K$ is the number of Kraus operators. Because this evaluation involves only matrix operations, it can be efficiently parallelized on GPUs. In practice with GPU acceleration, a single fidelity evaluation requires only a few seconds, yielding a speedup of one to two orders of magnitude speedup compared with conventional convex-optimization approaches~\cite{Reimpell2005IterativeOpt,reimpell2005optimal,Fletcher2007OptRecovery,watrous2009semidefinite,kosut2008robust,Audenaert2002OptSDP1,Mironowicz2024SDPandQI}. This advantage becomes particularly significant at large mean photon number $\bar{n}$ and in applications involving extensive parameter sweeps or learning-based opt	on, where fidelity evaluations must be performed repeatedly.

The near-optimal fidelity~\cite{Zheng2024NearOptimal} is achievable by the transpose channel by construction, and it is defined as 
\begin{equation}
    \tilde{F}^{\mathrm{opt}} = \frac{1}{d_L^2} \left\lVert \mathrm{Tr}_L \sqrt{M} \right\rVert^2 _F,
\end{equation}
where $M$ is the QEC matrix, $(\mathrm{Tr}_L B)_{l,k}=\sum_{\mu}B_{[\mu l],[\mu k]}$ denotes the partial trace over the code space indices, $\Vert \cdot \Vert_F$ is the Frobenius norm, and $d_L$ is the encoding dimension, which is two, since the QEC codes have two logical codewords. It satisfies the rigorous two-sided bound:
\begin{equation}\label{eq:infidelity_condition}
    \frac{1 - \tilde{F}^{\mathrm{opt}}}{2} \leq 1 - F^{\mathrm{opt}} \leq 1 - \tilde{F}^{\mathrm{opt}}.
\end{equation}
The width of this bound scales linearly with the optimal infidelity.
Therefore, when the code performs well, the near-optimal fidelity provides a quantitatively accurate approximation of the true optimal fidelity, and our results also illustrate the two-sided bound. Crucially, near-optimal fidelity can faithfully capture the dependence of error-correction performance on code parameters and noise strength.

Regarding numerical stability, matrix-based computations inevitably introduce numerical errors. We carefully benchmarked the accumulated numerical error at each stage of the calculation and verified that it remains at least two orders of magnitude smaller than the infidelity bound in Eq.~\eqref{eq:infidelity_condition}. Detailed error estimates are provided in Appendix~\ref{sec:error_analysis and Kraus}.

All simulations are implemented by \texttt{PyTorch}~\cite{Pytorch2019}, leveraging GPU acceleration and optimized linear-algebra operations. The primary resource limitation arises from the GPU memory, which scales with the truncated dimension and the number of Kraus operators. The peak memory consumption associated with the QEC matrix can be estimated as 
\begin{equation*}
\mathrm{Memory_{peak}} = \alpha \cdot (d_L N_K)^2 \times \rm{bytes\_per\_element},
\end{equation*}
where $\alpha$ is the peak memory factor and $N_K=N_{K_{\gamma t}}\cdot N_{K_{\kappa t}}$. Meanwhile, $N_K$ is positively correlated with the truncated dimension, and the details can be found in Appendix~\ref{sec:error_analysis and Kraus}. In our results, the estimated infidelity generally remains at a small magnitude, thus a very large number of Kraus operators is necessary. In practical applications, calculation parameters can be adjusted according to the real precision requirements. When necessary, computations can be performed on CPUs to bypass GPU memory constraints, enabling exploration of higher excitation numbers and stronger noise channels at the cost of acceleration.

% ---------------------------------------------------------------------------

\subsection{Optimizing code parameters and near-optimal fidelity with covariance matrix adaptation evolution strategy}
\label{sec:CMA-ES}

The task of maximizing the near-optimal fidelity under a given loss-dephasing channel amounts to a high-dimensional, nonconvex black-box optimization problem. The parameter landscape is generally multimodal, exhibits strong parameter coupling, and may contain both continuous and discrete degrees of freedom. In such settings, gradient-based methods tend to become trapped in local optima, while exhaustive grid searches are computationally prohibitive.

To address this challenge, we employ evolutionary algorithms~\cite{holland1975adaptation,goldberg1989genetic,schwefel1995evolution,back1996evolutionary,Dasgupta1997}, which are well suited for global search in complex, derivative-free landscapes. In our framework, the code parameters serve as the genotype, while the near-optimal fidelity plays the role of the fitness function. 

Among evolutionary algorithms, the covariance matrix adaptation evolution strategy (CMA-ES)~\cite{hansen2001completely,hansen2003reducing,Hansen2011CMAESES,hansen2014cmaes,2024CMAES} is a widely used derivative-free optimization method. It searches for the global optimum of an objective function by iteratively updating a multivariate Gaussian sampling distribution. The blue panel of Fig.~\ref{fig:workflow} illustrates the main principles of CMA-ES. 

At each generation, a population of candidate solutions is sampled around the current mean vector $m$ according to a covariance matrix $C$ and a global step-size $\sigma$. The candidates are evaluated through the fidelity objective, ranked according to their performance, and recombined to form an updated mean via weighted averaging.

A key feature of CMA-ES is the self-adaptation of both the covariance matrix and the step size. The covariance matrix is updated by accumulating information from successful search directions, thereby reinforcing promising directions in the parameter space. In parallel, the step size is adjusted according to the length of the evolution path, which controls the overall scale of exploration. This mechanism enables automatic balancing between global exploration and local refinement and, in practice, requires only two user-tunable hyperparameters (initial step size and population size), greatly simplifying its application.

For our simulations, we set the initial step size to $\sigma=0.3$ and the population size to 50. We observe substantial robustness with respect to these choices. Because CMA-ES is asymptotic in nature, convergence behavior depends on the complexity of the parameter space. To assess reliability, we performed repeated optimization runs. 

For continuous parameter families such as GKP codes, the results remain virtually unchanged. For parameter spaces involving discrete variables, as in NP codes, small fluctuations may appear; however, the resulting fidelity variations remain two orders of magnitude smaller than the intrinsic infidelity bound of the near-optimal fidelity metric. This confirms the numerical stability and credibility of our optimization results.

% ---------------------------------------------------------------------------
\begin{figure*}
    \centering
    \includegraphics[width=1\linewidth]{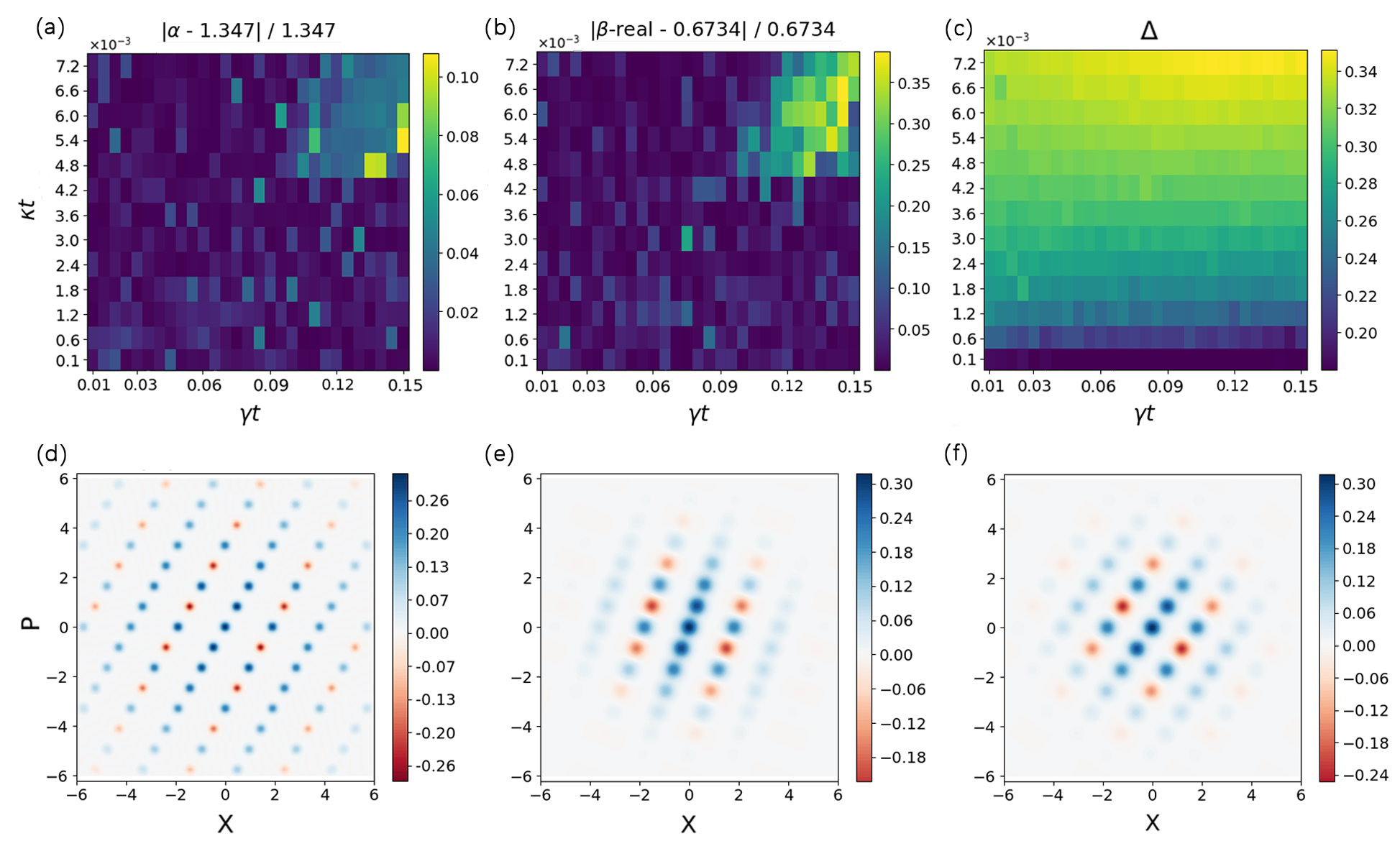}
    \caption{Optimal parameters of GKP codes vs loss and dephasing: (a) $\alpha$, (b) $\beta\mathrm{-real}$, and (c) $\Delta$. The shape parameters $\alpha$ and $\beta\mathrm{-real}$ are compared with the hexagonal lattice values ($\alpha \approx 1.347$ and $\beta\mathrm{-real} \approx 0.6734$). For most noise conditions, the optimized parameters remain close to the hexagonal lattice; systematic deviations appear when both loss and dephasing are strong. (d)-(f) show the Wigner functions of $|0_L\rangle$ for the hexagonal lattice and two deviated configurations. 
    The corresponding noise strengths and code parameters are: (d) $\{\gamma t=0.05, \kappa t=10^{-4}, \alpha=1.35, \beta=0.672+1.16 \mathrm{i}, \Delta=0.18\}$; (e) $\{\gamma t=0.15, \kappa t=7.2 \times10^{-3}, \alpha=1.29, \beta=0.443+1.22 \mathrm{i}, \Delta=0.35\}$; (f) $\{\gamma t=0.15, \kappa t=6.6 \times10^{-3}, \alpha=1.29, \beta=0.875+1.22 \mathrm{i}, \Delta=0.34\}$.
    }
    \label{fig:gkp_params}
\end{figure*}
% ---------------------------------------------------------------------------

The algorithm is implemented using the widely-used Python package \texttt{pycma}~\cite{hansen2019pycma}. Taken together, this framework enables efficient and reliable identification of near-globally optimal bosonic code parameters under general loss-dephasing noise. 
% ---------------------------------------------------------------------------
% ---------------------------------------------------------------------------

\section{Numerical Results and Discussion}
\label{sec:results}

In order to identify the performance boundary between GKP and NP codes under combined loss-dephasing noise, we systematically explored the parameter regime
\begin{equation}
    \gamma t \in[0.01,0.15], \quad \kappa t\in [0.0001, 0.0072].
\end{equation}

The sampling points on the $x$ axis (loss) are uniformly spaced at intervals of 0.005. The intervals on the $y$ axis (dephasing) are not uniform. Near the GKP advantage boundary in Fig.~\ref{fig:boundary}, we employ denser noise intervals (interval of 0.0001 when $\kappa t \in [0.0001, 0.0012]$) to make the boundary clear. We select 0.0001, 0.0006, and 0.0072 as representative points, and other sampling point results are not shown in the optimal parameter results because they exhibit similar patterns. And within the range of $[0.0012,0.0072]$, the interval is 0.0006. 

All simulations use double precision (\texttt{float64} and \texttt{complex128}). Truncation dimension and Kraus operator counts are carefully chosen to guarantee numerical convergence (see Appendixes~\ref{sec:truncation} and~\ref{sec:error_analysis and Kraus}). Kraus truncation is analytically estimated using Poisson or binomial distributions. 

% ---------------------------------------------------------------------------

\subsection{Optimal parameters of the GKP codes}
\label{sec:gkp_results}

As discussed in Sec.~\ref{sec:gkp_codes}, the shape parameters $\alpha$ and $\beta$ of the GKP lattice can in principle be chosen arbitrarily subject to the constraint in Eq.~\eqref{eq:GKP_condition}. In this optimization, the algorithm explores the full admissible parameter space. To reduce the dimensionality of the search, we fix $\alpha$ to lie along a lattice axis and therefore restrict $\alpha$ to be real. This choice does not compromise generality, since equivalent parametrizations are commonly adopted for square, rectangular, and hexagonal lattices. Together with Eq.~\eqref{eq:GKP_condition}, the constraint $\mathrm{Im}(\alpha) = 0$ reduces the independent variables to $\alpha$ and $\beta\mathrm{-real}$.

The unit-cell area of the GKP lattice is $\pi/2$. Taking lattice symmetries into account, we consider the parameter ranges $\alpha \in [0.1,\sqrt{\pi}]$ and $\beta\mathrm{-real} \in [-\sqrt{\pi},\sqrt{\pi}]$, which cover all inequivalent lattice configurations. For the energy parameter, we choose $\Delta \in [0.18,0.6]$. At $\Delta = 0.18$, the mean photon number is approximately $\bar{n} \approx 14.9$. Smaller values of $\Delta$ correspond to higher mean photon numbers but require substantially larger Fock-space truncations, leading to prohibitive GPU memory requirements. The full optimization domain is therefore
\begin{equation}
    \alpha\in[0.1,\sqrt{\pi}], \beta\mathrm{-real}\in[-\sqrt{\pi},\sqrt{\pi}], \Delta\in[0.18,0.6].
\end{equation}

Figures~\ref{fig:gkp_params} (a) and (b) show that the optimized values of $\alpha$ and $\beta\mathrm{-real}$ lie close to those of the hexagonal lattice, characterized by $\alpha \approx 1.347$ and $\beta\mathrm{-real} \approx 0.6734$. We plot the absolute value of $\beta\mathrm{-real}$, since opposite signs correspond to symmetry-related lattices and therefore equivalent excitation directions. 
Under most noise conditions, the optimized parameters are close to the hexagonal values, with relative deviations typically below $5\%$.

Furthermore, the fluctuations of $\beta\mathrm{-real}$ are significantly larger than those of $\alpha$, indicating that under combined loss and dephasing noise the GKP code primarily adapts through rotations of the lattice excitation direction rather than changes of the principal lattice axis. When both loss and dephasing become strong, systematic deviations from the hexagonal configuration appear. Multiple independent optimization runs yield consistent results, excluding stochastic optimization effects. Increasing the Fock-space truncation dimension and repeating the calculations on the CPU also leaves the deviations unchanged, ruling out truncation artifacts. These characteristics indicate that the hexagonal lattices are no longer strictly optimal in the regime of strong simultaneous loss and dephasing.

Figures~\ref{fig:gkp_params} (e) and (f) present the Wigner functions for two representative examples. The lattice preserves translational symmetry and retains an oblique structure, but exhibits a slight rotation of the excitation direction. This deformation appears to be a distinctive response of the GKP code to combined loss-dephasing noise, in contrast to the pure-loss regime.

Figure~\ref{fig:gkp_params} (c) shows the behavior of the energy parameter $\Delta$. As the dephasing strength increases, the optimal $\Delta$ increases correspondingly, implying a reduction in the mean photon number. For fixed dephasing, $\Delta$ depends only weakly on the loss strength. This behavior reflects the distinct physical mechanisms of the two noise channels. Dephasing acts directly in the photon-number basis and therefore strongly penalizes states with large mean photon number. Increasing $\Delta$ broadens the lattice peaks and suppresses number-dependent phase randomization. In contrast, photon loss primarily manifests as effective displacement noise in phase space, whose impact on logical information is largely governed by the lattice geometry rather than the peak width. As a result, the optimal $\Delta$ is much more sensitive to dephasing than to loss.

Overall, the optimized parameters show that the GKP code responds to combined loss and dephasing noise through coordinated adjustments of both lattice geometry and mean photon number. In the weak-dephasing regime ($\kappa t = 0.0001$), the optimization saturates the lower bound of $\Delta$, corresponding to the maximum allowed mean photon number. This behavior indicates that, in near-pure-loss channels, increasing the photon number monotonically improves error-correction performance. All observed trends are consistent with previous analyses.

In conclusion, GKP codes remain near the hexagonal-lattice configuration across most of the explored loss-dephasing parameter regime, with noticeable deviations occurring only when both noise channels become strong. Meanwhile, the optimal $\Delta$ increases primarily with dephasing strength and depends only weakly on photon loss. The optimized results therefore reveal a nontrivial trade-off between protection against loss and dephasing, mediated through the combined tuning of code energy and lattice geometry.

% ---------------------------------------------------------------------------

\subsection{Optimal parameters of the NP codes}
\label{sec:np_results}

The parameters of NP codes impose constraints on the required Fock-space truncation dimension (technical details are provided in the Appendix~\ref{sec:truncation}). As in any numerical simulation, only a finite parameter domain can be explored. In the present work, we consider
\begin{equation}
    \begin{aligned}
        &f\in [0.0,1.0], \quad s=[1,2,3,4,5],\\
        &r\in[-0.4,0.4], \quad n\in [1.0,4.0].
    \end{aligned}
\end{equation}
% and the algorithm can freely select parameters from it.

% ---------------------------------------------------------------------------
\begin{figure*}
    \centering
    \includegraphics[width=1.0\linewidth]{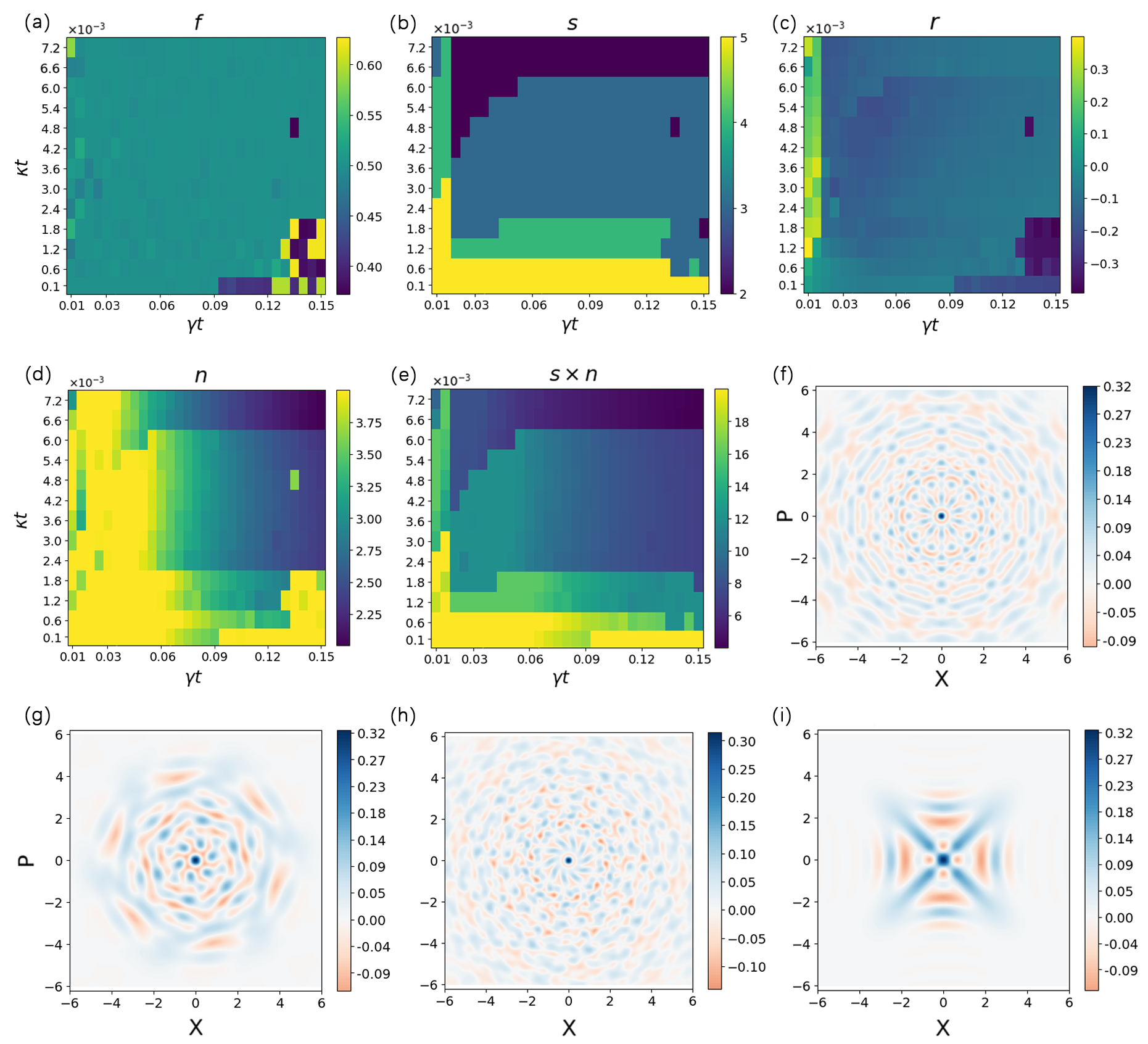}
    \caption{Optimal parameters of NP codes vs loss and dephasing: (a) $f$, (b) $s$, (c) Gaussian-state squeezing parameter $r$, (d) Gaussian-state mean photon number $n$, and (e) logical-state mean photon number $s \times n$. The parameters $f$ and $s$ determine the code-lattice geometry in the code space. (f)-(i) show the Wigner functions of $|0_L\rangle$ for four noise regimes: (f) lower left, (g) lower right, (h) top left, and (i) top right. 
    % The corresponding noise strengths and code parameters are indicated in the panel titles.
    The corresponding noise strengths and code parameters are: (f) $\{\gamma t=0.01, \kappa t=10^{-4}, f=0.500, s=5, r=0.0527, n=4.00\}$; (g) $\{\gamma t=0.01, \kappa t=7.2\times10^{-3}, f=0.584, s=3, r=0.3245, n=3.05\}$; (h) $\{\gamma t=0.15, \kappa t=10^{-4}, f=0.600, s=5, r=0.2309, n=4.00\}$; (i) $\{\gamma t=0.15, \kappa t=7.2\times10^{-3}, f=0.500, s=2, r=-0.0720, n=2.01\}$.
    }
    \label{fig:np_params}
\end{figure*}
% ---------------------------------------------------------------------------

In the simulations, we explicitly construct the logical states $|0\rangle_L$ and $|1\rangle_L$. Figure~\ref{fig:np_params} summarizes the optimized NP code parameters as functions of the loss and dephasing strengths. Compared with the GKP codes, the NP parameter space is structurally more complex, primarily due to the presence of the discrete parameter $s$, which leads to more pronounced fluctuations in the optimized parameters. Nevertheless, repeated optimization runs confirm that the associated fidelity variations remain small: the resulting fidelity differences are at least two orders of magnitude smaller than the intrinsic infidelity bound of the near-optimal fidelity, ensuring that these fluctuations do not affect the qualitative conclusions.

The parameters $s$ and $f$ determine the lattice geometry of the code space. In the previous work~\cite{Hu2025NP}, we showed that $f = 1/2$ corresponds to the diamond lattice, which possesses the highest symmetry. For this lattice, the code distances are $d_N = 2s$ in the number direction and $d_{\phi} = \pi/s$ in the phase direction. Here, our numerical results show that when photon loss dominates, the optimization favors a larger $d_N$, whereas strong dephasing favors a larger $d_{\phi}$. In most noise regimes, the optimal value of $f$ remains close to $1/2$, consistent with the preference for highly symmetric lattices.

However, in the regime of high-loss and weak-dephasing (notably in the lower right region of Fig.~\ref{fig:np_params}), the optimal parameter occasionally shifts to $f=2/5$ or $3/5$, accompanied by a reduction in $s$. This behavior can be understood in terms of the effective code distance in the number direction, which scales as $d_N \cdot \mathrm{den}(f)$, where $\mathrm{den}(f)$ denotes the denominator of the rational parameter $f$. Changing $f$ from $1/2$ to $3/5$ increases the denominator from 2 to 5 and therefore enhances the effective distance against excitation loss. Because the maximal value of $s$ is restricted in the explored parameter region, the optimization compensates by adjusting $f$ to increase the effective distance along the loss direction. We therefore expect that if larger values of $s$ were accessible, the optimal solution would revert to the symmetric case $f = 1/2$.

The squeezing parameter $r$ exhibits a clear dependence on the loss strength. When the loss is substantial, the optimal $r$ consistently takes negative values, whereas positive values occur only at small loss strengths ($\gamma t = 0.01$ and $0.015$). Negative $r$ reduces the phase uncertainty of the Gaussian envelope and thereby improves distinguishability of phase-direction errors. In the low-loss regime, the NP code favors Gaussian states with larger mean photon number, for which the phase uncertainty is already sufficiently small; in this case, positive values of $r$ help suppress errors associated with higher-order photon loss events.

The Gaussian-state photon number $n$ and the logical-state photon number $s \times n$ also reflect a trade-off between loss and dephasing. In general, increasing noise strength leads to lower optimal energy. The selected energy depends not only on the total noise magnitude but also on the relative dominance of loss and dephasing, as well as on the lattice geometry. When one noise mechanism dominates, the optimization favors energy configurations that enhance protection against that channel. In particular, in the regimes where $f = 2/5$ or $3/5$, the parameter $r$ tends to assume more negative values while the Gaussian state adopts a larger mean photon number, indicating a coordinated adjustment of lattice geometry and state squeezing to counteract strong photon loss.

Overall, the optimized NP codes generally favor lattice geometries close to the diamond lattice ($f \approx 1/2$), while $f=2/5$ or $3/5$ becomes favorable only in the high-loss, weak-dephasing regime due to their larger effective code distance against excitation loss. Meanwhile, the energy-related parameters depend jointly on the lattice geometry and the dominant noise mechanism. As a result, the optimal logical-state energy is governed by the competition between loss and dephasing, rather than increasing monotonically with the mean photon number.

% ---------------------------------------------------------------------------

\subsection{Performances and the performance boundary of two codes}
\label{sec:comparison} 

\begin{figure}
    \centering
    \includegraphics[width=1\linewidth]{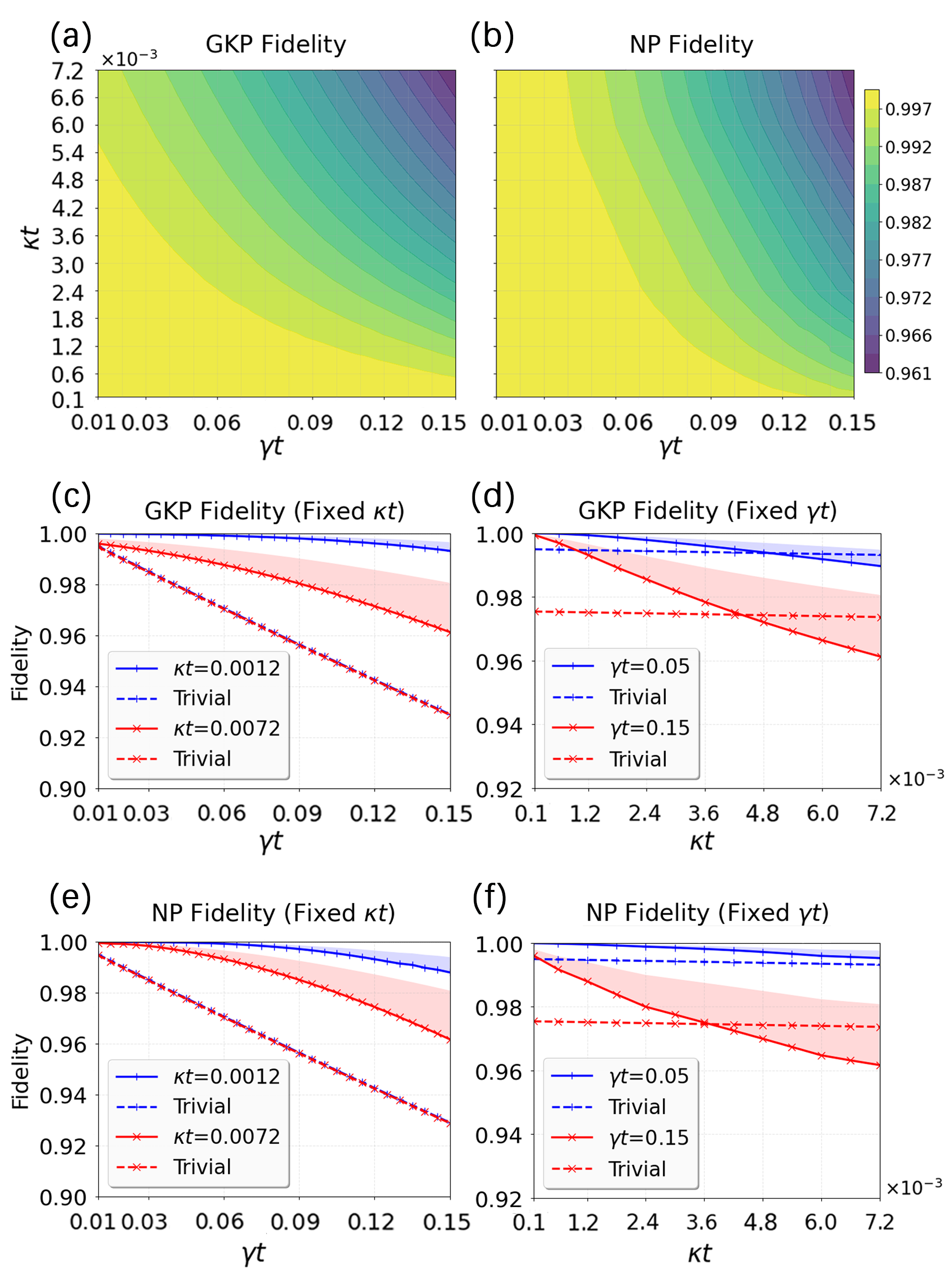}
    \caption{(a), (b) Contour maps of near-optimal fidelity: (a) $\tilde{F}^{\mathrm{opt}}_{\mathrm{GKP}}$ and (b) $\tilde{F}^{\mathrm{opt}}_{\mathrm{NP}}$ vs loss and dephasing.  (c)-(f) Corresponding bounds on the optimal fidelity, and the dashed lines are estimated by the trivial encoding using Fock states $|0\rangle$ and $|1\rangle$. (c), (e) Fidelity vs photon loss for fixed dephasing $\kappa t =0.0012$ (blue) and $0.0072$ (red). (d), (f) Fidelity vs dephasing for fixed photon loss $\gamma t=0.05$ (blue) and $0.15$ (red). Shaded regions indicate the range of optimal fidelity consistent with the two-sided bound inferred from the near-optimal fidelity.}
    \label{fig:performances}
\end{figure}

Figure~\ref{fig:performances} presents a performance comparison between GKP and NP codes. Figs.~\ref{fig:performances} (a) and (b) show the near-optimal fidelity as a function of photon loss and dephasing strength, while Figs.~\ref{fig:performances} (c)-(f) display the corresponding two-sided bounds on the optimal fidelity, and the dashed lines are estimated by the trivial encoding using Fock states $|0\rangle$ and $|1\rangle$.

The fidelity curves exhibit distinct shapes along the loss and dephasing directions. As shown in Figs.~\ref{fig:performances} (c)-(f), the fidelity is concave as a function of photon loss but convex as a function of dephasing. This difference originates from the distinct scaling of the two noise mechanisms. To leading order, photon loss and dephasing are generated by the operators $\sqrt{\gamma t}\hat{a}$ and $\sqrt{\kappa t}\hat{a}^\dagger \hat{a}$, respectively. Consequently, the fidelity degradation induced by dephasing effectively exhibits a square-root scaling relative to that of photon loss. After rescaling the dephasing axis by a square-root transformation, the resulting curves recover a concave profile similar to that of the loss direction.

% ---------------------------------------------------------------------------
% ---------------------------------------------------------------------------

A direct comparison of Figs.~\ref{fig:performances} (a) and (b) reveals a clear division of advantageous regimes. GKP codes outperform NP codes in the photon-loss-dominated noise channels, whereas NP codes exhibit superior robustness against dephasing. Moreover, GKP codes are substantially more sensitive to dephasing than NP codes are to photon loss. In the regime where both noise channels are weak, NP codes typically achieve higher fidelity. These trends imply the existence of a performance boundary in the joint loss-dephasing parameter space. This boundary delineates the relative advantages of the two bosonic lattice codes and provides quantitative guidance for selecting an encoding under a given combined noise model.

% ---------------------------------------------------------------------------
\begin{figure}
    \centering
    \includegraphics[width=1\linewidth]{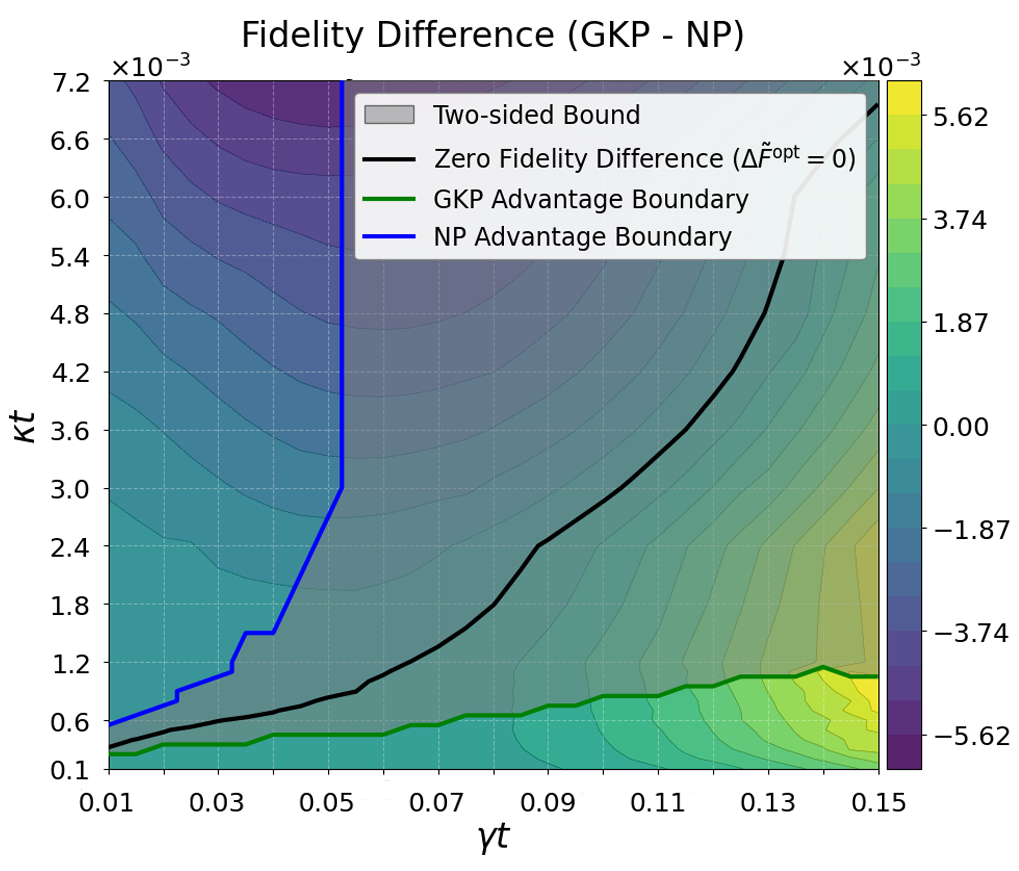}
    \caption{Contour map of the near-optimal fidelity difference $\Delta \tilde{F} = \tilde{F}^{\mathrm{opt}}_{\mathrm{GKP}} - \tilde{F}^{\mathrm{opt}}_{\mathrm{NP}}$ as a function of photon loss and dephasing. The green and blue curves denote the strict advantage boundaries of the GKP and NP codes, respectively. The black curve indicates the contour $\Delta\tilde{F}^{\mathrm{opt}}=0$.}
    \label{fig:boundary}
\end{figure}
% ---------------------------------------------------------------------------

To characterize this boundary, we evaluate the fidelity difference 
\begin{equation*}
\Delta \tilde{F}^{\mathrm{opt}} = \tilde{F}_{\mathrm{GKP}}^{\mathrm{opt}} - \tilde{F}_{\mathrm{NP}}^{\mathrm{opt}},
\end{equation*}
and plot the resulting contour map in Fig.~\ref{fig:boundary}, using linear interpolation between sampled points. From Eq.~\eqref{eq:infidelity_condition}, the optimal fidelity satisfies the bounds as \begin{equation*}
\tilde{F}^{\mathrm{opt}} \le F^{\mathrm{opt}} \le \frac{1+\tilde{F}^{\mathrm{opt}}}{2}.
\end{equation*}
We define a strict GKP advantage regime by
\begin{equation*}
    F^{\mathrm{GKP}}_{\rm{lower}} > F^{\mathrm{NP}}_{\rm{upper}},
\end{equation*}
and analogously define the NP advantage regime by 
\begin{equation*}
F^{\mathrm{NP}}_{\rm{lower}} > F^{\mathrm{GKP}}_{\rm{upper}}.
\end{equation*}
The resulting boundaries are shown as the green and blue boundaries in Fig.~\ref{fig:boundary}. This definition ensures the identified regions correspond to an unambiguous performance advantage. The black curve indicates the contour $\Delta \tilde{F}^{\mathrm{opt}}=0$, where the two codes exhibit equal near-optimal fidelity.

The performance boundary is reached when the dephasing strength is approximately two orders of magnitude smaller than the photon-loss strength.
Within the explored parameter regime
\begin{equation*}
    \gamma t \in [0.01,0.15], \quad
\kappa t \in [10^{-4},7.2\times10^{-3}],
\end{equation*}
the boundary emerges approximately along the scaling relation
\begin{equation}
    \kappa t \sim 10^{-2}\gamma t.
\end{equation}
Along this boundary, the optimized near-optimal fidelities of the two code families become comparable.
Beyond this boundary, higher fidelity can be achieved by selecting the code within its respective advantageous regime.

This result also hints at the possibility that, in the vicinity of the boundary, alternative bosonic codes capable of outperforming both GKP and NP codes may exist.
Finding the optimal bosonic code for a given loss-dephasing channel remains a long-standing open problem. Previous studies have shown that bosonic codes approaching the maximum channel capacity often exhibit hybrid features of both GKP and NP codes~\cite{Leviant2022quantumcapacity}.
In that work, bosonic codes numerically optimized for loss-dephasing channels via biconvex optimization were found to resemble hexagonal GKP states in the pure-loss limit and cat-like states in the pure-dephasing limit. Under combined loss-dephasing channels, the Wigner functions of the optimized codes exhibit a hybrid structure combining translational and rotational symmetries.
Although such numerically optimized states are difficult to realize experimentally, they reveal intrinsic structural properties of bosonic modes under different noise mechanisms.

In contrast, the present work focuses on optimizing codes constrained to possess either translational symmetry (GKP codes) or rotational symmetry (NP codes). The resulting performance boundary, therefore, serves not only as a practical guideline for code selection but also as an indirect indicator of the regime in which such a symmetry transition is expected to occur.

In addition, we clarify that the optimization domains for the GKP and NP code families were chosen to ensure numerical feasibility and convergence, rather than to impose an exactly identical upper bound on the mean photon number. For approximate GKP codes, the mean photon number can be estimated as~\cite{gh1c-xyn1}
\begin{equation}
    \bar n_{\mathrm{GKP}}
\simeq
\frac{1}{e^{2\Delta^2}-1},
\end{equation}
so that the parameter range explored in our calculations corresponds approximately to $\bar n_{\mathrm{GKP}}\lesssim 14.9$. For NP codes, the optimized parameter region typically gives logical-state photon numbers $\bar n_{\mathrm{NP}} = s\times n \lesssim 20$. Thus, the two code families are compared within similar energy scales, although not under a strictly identical energy cutoff.

We emphasize that this difference in the accessible energy range does not affect the main physical conclusions. Taken together, Figs.~2(c), 3(e), and 5 show that GKP codes exhibit a clear advantage in the loss-dominated regime even though the optimized NP codes often possess larger average photon numbers. This indicates that the observed boundary is not primarily an artifact of the energy window, but instead reflects the distinct symmetry structures and error sensitivities of two code families: translationally symmetric GKP codes are naturally suited to loss-dominated displacement-type errors, whereas rotationally symmetric NP codes become more favorable as dephasing becomes increasingly relevant.

Moreover, the optimized fidelities are generally not obtained at the largest allowed code energies. Except in the extremely weak-dephasing regime, both code families typically achieve optimal performance at intermediate energies, determined by the competition between loss and dephasing. This behavior reflects the nontrivial trade-off between increasing code distance and the heightened noise sensitivity associated with higher photon numbers. Therefore, although the two optimization domains are not defined by an identical energy cutoff, they remain sufficiently broad and comparable in energy scale to reliably identify the performance boundary between the optimized GKP and NP code families under the loss-dephasing channels considered here.

\section{Summary}
\label{sec:summary}

So far, we have developed a scalable and systematic code-parameter optimization framework that maximizes the performance of bosonic codes under a general loss-dephasing noise model, exemplified here by the GKP and NP codes. In practice, optimal code parameters can be efficiently obtained by specifying the loss and dephasing strengths and tuning only two hyperparameters of the CMA-ES optimizer, making the method straightforward to deploy across a wide range of noise conditions. While our numerical demonstrations focus on a finite window of loss-dephasing strengths to observe the performance boundary between code families, the approach itself is readily extensible to broader noise ranges and higher code energies, subject only to computational resources. 

Using this method, we perform a quantitative comparison of the GKP and NP codes and numerically identify a performance boundary separating the noise regime in which each code exhibits a fundamental advantage. We find that the boundary emerges when the dephasing strength is approximately two orders of magnitude smaller than the loss strength. Furthermore, the optimal code parameters exhibit interpretable trends as noise strength varies, reflecting the underlying trade-off between displacement and rotation error suppression. Our results delineate the regimes in which the GKP and NP codes achieve an absolute performance advantage and suggest a possible bosonic encoding with superior performance on this boundary.

In summary, our work establishes a practical methodology for quantitatively benchmarking bosonic QEC codes and exploring their performance landscapes in realistic noise environments. As such, it provides concrete guidance for the experimental selection of both code families and optimized parameters, and opens the door to systematic numerical studies of more bosonic encodings and noise models.
% ---------------------------------------------------------------------------
% ---------------------------------------------------------------------------

\begin{acknowledgments}
This work is supported by the National Natural Science Foundation of China (Grant No. 12375025).
\end{acknowledgments}

% ---------------------------------------------------------------------------
% ---------------------------------------------------------------------------

\appendix

\section{Kraus operator representation of the loss-dephasing channel}
\label{sec:kraus_ope}

To consider a noise channel consisting of simultaneous photon loss and dephasing, the Lindblad master equation can be described as
\begin{equation}
    \frac{\partial \hat{\rho}(t)}{\partial t} = \gamma \mathcal{L}[\hat{a}]\hat{\rho}(t) + \kappa \mathcal{L}[\hat{n}]\hat{\rho}(t),
\end{equation}
with $\mathcal{L}[\hat{A}]\hat{\rho}(t)=\hat{A}\hat{\rho}(t)\hat{A}^{\dagger}-1/2 \rho(t)\hat{A}^{\dagger}\hat{A}-1/2\hat{A}^{\dagger}\hat{A}\rho(t)$.
The first term describes photon loss at a rate $\gamma$, while the second term accounts for pure dephasing at rate $\kappa$.
By integrating it over time, we can derive the noise channels $\mathcal{N}(\hat{\rho})$.

The Kraus operator decomposition of $\mathcal{N}(\hat{\rho})$ can be written as
\begin{equation}
    \mathcal{N}(\hat{\rho}) = \int^{\infty}_{-\infty} d\phi P(\phi) \sum^{\infty}_{k=0} \hat{R}(\phi)\hat{A}_k \hat{\rho} \hat{A}_k^\dagger \hat{R}(\phi)^\dagger,
    \label{eq:N_form1}
\end{equation}
where $\hat{A}_k$ is given by
\begin{equation}\label{eq:A_K}
    \hat{A}_k = \frac{1}{\sqrt{k!}} (1-e^{-\gamma t})^{k/2} e^{-\gamma t\hat{n}/2} \hat{a}^k,
\end{equation}
which is related to the loss. Meanwhile, $P(\phi) = (2\pi \kappa t)^{-1/2}e^{-\phi^2 / (2\kappa t)}$ is a Gaussian with zero mean and variance $\kappa t$, and $\hat{R}(\phi) = e^{i\phi \hat{n}}$ is the phase rotation operator. We can also find an equivalent Kraus operator representation of the dephasing term as
\begin{equation}
    \sum^{\infty}_{l=0} \hat{B}_l \hat{\rho}\hat{B}^\dagger_l,
\end{equation}
with
\begin{equation}
    \hat{B}_l = \frac{(\kappa t)^{l/2}}{\sqrt{l!}}e^{-\kappa t \hat{n}^2 /2}\hat{n}^l.
\end{equation}
So we can rewrite Eq.~\eqref{eq:N_form1} as follows:
\begin{equation}
    \mathcal{N}(\hat{\rho}) = \sum_{k,l\geq0}\hat{B}_l \hat{A}_k \hat{\rho} \hat{A}_k^\dagger \hat{B}_l^\dagger.
    \label{eq:N_form2}
\end{equation}
% ---------------------------------------------------------------------------
% ---------------------------------------------------------------------------

\section{Hilbert space truncation}
\label{sec:truncation}

In numerical computation, we need to truncate the infinite-dimensional Hilbert space to finite-dimensional matrices. Selecting the truncated dimension $N$ for quantum states or operators is equivalent to ensuring that the discarded tail probability (or energy) remains below an acceptable error threshold $\varepsilon$. Let $\Pi_N = \sum^{N-1}_{n=0} |n\rangle\langle n|$ denote the projection operators of the truncated Fock basis $\mathcal{H}_N = \text{span}\{|0\rangle, |1\rangle, \dots, |N-1\rangle\}$, and for any state $\rho$, we define the weight of the population tail as
\begin{equation}
    \varepsilon_N = \mathrm{tr}[\rho(1-\Pi_N)] = \sum_{n \geq N}\langle n|\rho|n\rangle.
\end{equation}
For a pure state $|\psi\rangle = \sum_{n=0}^{\infty} c_n |n\rangle$, the truncated state is $|\psi_N\rangle = \Pi_N |\psi \rangle/\sqrt{1-\varepsilon_N}$, so the fidelity of the two states is $F = |\langle \psi | \psi_N \rangle|^2 = 1 - \varepsilon_N$. Therefore, the truncation error $\varepsilon_N$ directly describes the effect of truncation.

The approximate GKP states are defined by an envelope operator $e^{-\Delta^2 \hat{n}}$ acting on an infinite superposition. The probability of finding $n$ photons is the square of the amplitude:
\begin{equation}
    P(n) = |\langle n | \tilde{\mu} \rangle|^2  \approx e^{-2\Delta^2 n}.
\end{equation}
We require the truncation error $\varepsilon_{\rm{GKP}}$ to be less than the specified tolerance:
\begin{equation}
    \varepsilon_{\rm{GKP}} = \sum_{n=N_{\rm{cut}}}^{\infty} P(n) \approx \int_{N_{\rm{cut}}}^{\infty} e^{-2\Delta^2 n} dn \approx \frac{e^{-2\Delta^2 N_{\rm{cut}}}}{2\Delta^2} < \varepsilon_{\rm{tol}}.
\end{equation}
Neglecting the constant, now we have
\begin{equation}
    N_{\rm{cut}} \approx \frac{-\mathrm{ln}(\varepsilon_{\rm{tol}})}{2\Delta^2}.
\end{equation}

As for determining $\varepsilon_{\rm{tol}}$, we devise a dynamic estimation method based on the magnitude of fidelity corresponding to the noise, ensuring that $\varepsilon_{\rm{GKP}}$ is two orders of magnitude smaller than the infidelity. Moreover, we verify that the estimated truncated dimension is sufficiently large based on the convergence of the photon number of logical states.

The approximate NP states simulated in our experiment are the logical $|0\rangle_L$ and $|1\rangle_L$ states. The truncation of the Hilbert space relies on the mean number and variance of the photon distribution. For a nonsqueezed coherent state, its photon number variance is equal to its mean photon number $\bar{n}$. Therefore, we choose the Hilbert space cut $N_{\rm{cut}}=2s\bar{n}$ to ensure the completeness of the Hilbert space. However, when the squeezing parameter $r$ is nonzero, it needs to be corrected to 
\begin{equation}
    N_{\rm{cut}} \approx2s\textrm{Var}_{\textrm{code}}(\hat{n}).
\end{equation}
% ---------------------------------------------------------------------------
% ---------------------------------------------------------------------------

\section{Error analysis and estimation of the number of Kraus operators}
\label{sec:error_analysis and Kraus}

One often encounters numerous difficulties in numerical simulation of quantum systems, including truncation errors in discrete models, rounding errors due to finite-precision arithmetic, and numerical truncation errors arising from finite approximations of infinite processes \cite{Provazník2022Taming}. In our work, numerical errors primarily stem from three sources: the construction of quantum states, the construction of Kraus operators, and the operations on QEC matrices. In Appendix.~\ref{sec:truncation}, we have already discussed truncated error control concerning the construction of quantum states. 
To avoid the impact of rounding errors, we select \texttt{float64} and \texttt{complex128} as the data precision, employ stabilized algorithms at every step, and perform pre- and post-comparisons of the error introduced at each computational step to prevent cumulative error. In practice, the operations on QEC matrices do not introduce errors that could affect the results.

The construction of Kraus operators is influenced by both the truncated dimension and the number of terms. The truncated dimension corresponds to the quantum states. The number of terms affects the completeness relation of Kraus operators:
\begin{equation}
    \sum_i \hat{K}_i^\dagger \hat{K}_i = I.
\end{equation}
Since numerical computations must truncate infinite-dimensional operators and infinite series terms, the occurrence of errors is inevitable. We have to quantify the error introduced by these two truncations. 

The eigenvalues of the bosonic mode depend on the number of photons $n$, so subsequently the error can be expressed as
\begin{equation}
    \epsilon_{\mathrm{trunc}} = \max_{n \in [0, n_{\max}]} \left| 1 - \sum_{k=0}^{K_{\mathrm{cut}}} \langle n | \hat{K}_k^\dagger \hat{K}_k | n \rangle \right|,
\end{equation}
where $\hat{K}_k$ contain both loss and dephasing channels.

Our problem is to estimate the truncation indices $k_{\max}$ for Kraus operators $\hat{A}_k$ such that the approximate evolution
\begin{equation}
    \mathcal{N}(\rho) = \sum_{k=0}^{\infty} \hat{A}_k \rho \hat{A}_k^\dagger \approx \sum_{k=0}^{k_{\max}} \hat{A}_k \rho \hat{A}_k^\dagger
\end{equation}
satisfies the accuracy requirement:
\begin{equation}
    \varepsilon_{\mathrm{loss}} := \max_{\rho \in \mathcal{D}(\mathcal{H}_C)} \left[1 - \sum_{k=0}^{k_{\max}} \operatorname{Tr}\left(\hat{A}_k \rho \hat{A}_k^\dagger\right)\right] < \epsilon,
\end{equation}
where $\mathcal{H}_C$ is the code subspace. A similar requirement applies to $\hat{B}_l$ for the dephasing channel.

The photon loss channel has Kraus operators
\begin{equation}
    \hat{A}_k = \sqrt{\frac{(1 - \eta)^{k/2}}{k!}} \, \eta^{\hat{n}/2} \hat{a}^k
    \quad \text{with } \quad \eta = e^{-\gamma t},
\end{equation}
where each $\hat{A}_k$ represents the loss of $k$ photons. For a Fock state $|n\rangle$ with $n$ photons, the probability of losing exactly $k$ photons ($0 \leq k \leq n$) is given by
\begin{equation}
    p_k^{(n)} = \operatorname{Tr}[\hat{A}_k |n\rangle\langle n| \hat{A}_k^\dagger] = \binom{n}{k} (1-\eta)^k \eta^{n-k},
\end{equation}
which follows a binomial distribution with success probability $p = 1-\eta = 1-e^{-\gamma t}$.

To ensure that the truncation error remains below $\epsilon$ for all states in the code subspace, we consider the worst-case scenario in which the state contains the maximum number of photons $n_{\max}$ supported by the truncated Fock space. We then require
\begin{equation}
    \sum_{k=k_{\max}+1}^{n_{\max}} \binom{n_{\max}}{k} (1-\eta)^k \eta^{n_{\max}-k} < \epsilon,
\end{equation}
which can be expressed in terms of the binomial percent point function (ppf):
\begin{equation}
    k_{\max} = \operatorname{binomial.ppf}(1 - \epsilon, n_{\max}, 1 - e^{-\gamma t}).
\end{equation}

For the dephasing channel, the Kraus operators are
\begin{equation}
    \hat{B}_{l} = \frac{(\kappa t)^{l/2}}{\sqrt{l!}} e^{- \kappa t \hat{n}^2 /2} \hat{n}^{l}.
\end{equation}
For a Fock state $|n\rangle$, the probability associated with the $l$-th term is
\begin{equation}
    p_l^{(n)} = \operatorname{Tr}[\hat{B}_l |n\rangle\langle n| \hat{B}_l^\dagger] = \frac{(\kappa t n^2)^l}{l!} e^{-\kappa t n^2},
\end{equation}
which follows a Poisson distribution with parameter $\lambda_n = \kappa t n^2$. Taking the conservative bound $n = n_{\max}$, we require
\begin{equation}
    \sum_{l=l_{\max}+1}^{\infty} \frac{(\kappa t n_{\max}^2)^l}{l!} e^{-\kappa t n_{\max}^2} < \epsilon,
\end{equation}
leading to
\begin{equation}
    l_{\max} = \operatorname{poisson.ppf}(1 - \epsilon, \kappa t n_{\max}^2).
\end{equation}

The standard deviation of the probability distribution can be configured based on the GPU's memory size and the required precision. Now we can dynamically estimate the terms of the Kraus operators required based on $\gamma$ and $\kappa$. Moreover, to prevent the estimation of an excessively small number of Kraus operators during periods of low noise, we set a minimum value to ensure numerical stability. As such, we can control the error within the required precision.

\bibliography{reference}

@article{gottesman2001encoding,
  title = {Encoding a qubit in an oscillator},
  author = {Gottesman, Daniel and Kitaev, Alexei and Preskill, John},
  journal = {Phys. Rev. A},
  volume = {64},
  issue = {1},
  pages = {012310},
  numpages = {21},
  year = {2001},
  month = {Jun},
  publisher = {American Physical Society},
  doi = {10.1103/PhysRevA.64.012310},
  url = {https://link.aps.org/doi/10.1103/PhysRevA.64.012310}
}

@article{Grimsmo2021GKPCode,
  title = {Quantum Error Correction with the Gottesman-Kitaev-Preskill Code},
  author = {Grimsmo, Arne L. and Puri, Shruti},
  journal = {PRX Quantum},
  volume = {2},
  issue = {2},
  pages = {020101},
  numpages = {20},
  year = {2021},
  month = {Jun},
  publisher = {American Physical Society},
  doi = {10.1103/PRXQuantum.2.020101},
  url = {https://link.aps.org/doi/10.1103/PRXQuantum.2.020101}
}

@Article{Hu2025NP,
author={Hu, Dong-Long
and Cai, Weizhou
and Zou, Chang-Ling
and Xiang, Ze-Liang},
title={Generalized number-phase lattice encoding of a bosonic mode for quantum error correction},
journal={Nat. Commun.},
year={2025},
month={Aug},
day={16},
volume={16},
number={1},
pages={7647},
issn={2041-1723},
doi={10.1038/s41467-025-62898-1},
url={https://doi.org/10.1038/s41467-025-62898-1}
}

@article{Grimsmo2020RotationCodes,
  title = {Quantum Computing with Rotation-Symmetric Bosonic Codes},
  author = {Grimsmo, Arne L. and Combes, Joshua and Baragiola, Ben Q.},
  journal = {Phys. Rev. X},
  volume = {10},
  issue = {1},
  pages = {011058},
  numpages = {32},
  year = {2020},
  month = {Mar},
  publisher = {American Physical Society},
  doi = {10.1103/PhysRevX.10.011058},
  url = {https://link.aps.org/doi/10.1103/PhysRevX.10.011058}
}

@article{Zheng2024NearOptimal,
  title = {Near-Optimal Performance of Quantum Error Correction Codes},
  author = {Zheng, Guo and He, Wenhao and Lee, Gideon and Jiang, Liang},
  journal = {Phys. Rev. Lett.},
  volume = {132},
  issue = {25},
  pages = {250602},
  numpages = {6},
  year = {2024},
  month = {Jun},
  publisher = {American Physical Society},
  doi = {10.1103/PhysRevLett.132.250602},
  url = {https://link.aps.org/doi/10.1103/PhysRevLett.132.250602}
}

@article{Albert2018PerformanceSingleMode,
  title = {Performance and structure of single-mode bosonic codes},
  author = {Albert, Victor V. and Noh, Kyungjoo and Duivenvoorden, Kasper and Young, Dylan J. and Brierley, R. T. and Reinhold, Philip and Vuillot, Christophe and Li, Linshu and Shen, Chao and Girvin, S. M. and Terhal, Barbara M. and Jiang, Liang},
  journal = {Phys. Rev. A},
  volume = {97},
  issue = {3},
  pages = {032346},
  numpages = {30},
  year = {2018},
  month = {Mar},
  publisher = {American Physical Society},
  doi = {10.1103/PhysRevA.97.032346},
  url = {https://link.aps.org/doi/10.1103/PhysRevA.97.032346}
}

@article{Schlegel2022CatStates,
  title = {Quantum error correction using squeezed Schr\"odinger cat states},
  author = {Schlegel, David S. and Minganti, Fabrizio and Savona, Vincenzo},
  journal = {Phys. Rev. A},
  volume = {106},
  issue = {2},
  pages = {022431},
  numpages = {17},
  year = {2022},
  month = {Aug},
  publisher = {American Physical Society},
  doi = {10.1103/PhysRevA.106.022431},
  url = {https://link.aps.org/doi/10.1103/PhysRevA.106.022431}
}

@misc{totey2023performancerandombosonicrotation,
      title={The performance of random bosonic rotation codes}, 
      author={Saurabh Totey and Akira Kyle and Steven Liu and Pratik J. Barge and Noah Lordi and Joshua Combes},
      year={2023},
      eprint={2311.16089},
      archivePrefix={arXiv},
      primaryClass={quant-ph},
      url={https://arxiv.org/abs/2311.16089}, 
}

@incollection{Pytorch2019,
title = {PyTorch: An Imperative Style, High-Performance Deep Learning Library},
author = {Paszke, Adam and Gross, Sam and Massa, Francisco and Lerer, Adam and Bradbury, James and Chanan, Gregory and Killeen, Trevor and Lin, Zeming and Gimelshein, Natalia and Antiga, Luca and others},
booktitle = {Advances in Neural Information Processing Systems 32},
pages = {8024--8035},
year = {2019},
publisher = {Curran Associates, Inc.},
url = {http://papers.neurips.cc/paper/9015-pytorch-an-imperative-style-high-performance-deep-learning-library.pdf}
}

@article{hansen2001completely,
  title={Completely Derandomized Self-Adaptation in Evolution Strategies},
  author={Hansen, Nikolaus and Ostermeier, Andreas},
  journal={Evolutionary Computation},
  volume={9},
  number={2},
  pages={159--195},
  year={2001},
  publisher={MIT Press},
  doi={10.1162/106365601750190398}
}

@article{hansen2003reducing,
  title={Reducing the Time Complexity of the Derandomized Evolution Strategy with Covariance Matrix Adaptation {(CMA-ES)}},
  author={Hansen, Nikolaus},
  journal={Evolutionary Computation},
  volume={11},
  number={1},
  pages={1--18},
  year={2003},
  publisher={MIT Press},
  doi={10.1162/106365603321828970}
}

@inproceedings{Hansen2011CMAESES,
author = {Hansen, Nikolaus and Auger, Anne},
title = {CMA-ES: evolution strategies and covariance matrix adaptation},
year = {2011},
isbn = {9781450306904},
publisher = {Association for Computing Machinery},
address = {New York, NY, USA},
url = {https://doi.org/10.1145/2001858.2002123},
doi = {10.1145/2001858.2002123},
booktitle = {Proceedings of the 13th Annual Conference Companion on Genetic and Evolutionary Computation},
pages = {991–1010},
numpages = {20},
location = {Dublin, Ireland},
series = {GECCO '11}
}

@article{hansen2014cmaes,
  title={{CMA-ES} for Derivative-Free Optimization: A Review},
  author={Hansen, Nikolaus},
  journal={arXiv:1604.00772},
  year={2014},
  note={Preprint, arXiv},
  url={https://arxiv.org/abs/1604.00772}
}

@article{2024CMAES,
title = {Covariance matrix adaptation evolution strategy based on correlated evolution paths with application to reinforcement learning},
journal = {Expert Systems with Applications},
volume = {246},
pages = {123289},
year = {2024},
issn = {0957-4174},
doi = {https://doi.org/10.1016/j.eswa.2024.123289},
url = {https://www.sciencedirect.com/science/article/pii/S0957417424001544},
author = {Oladayo S. Ajani and Abhishek Kumar and Rammohan Mallipeddi}
}

@INPROCEEDINGS{9317892,
  author={Jiang, Haotian and Kathuria, Tarun and Lee, Yin Tat and Padmanabhan, Swati and Song, Zhao},
  booktitle={2020 IEEE 61st Annual Symposium on Foundations of Computer Science (FOCS)}, 
  title={A Faster Interior Point Method for Semidefinite Programming}, 
  year={2020},
  volume={},
  number={},
  pages={910-918},
  doi={10.1109/FOCS46700.2020.00089}}

@book{holland1975adaptation,
  title={Adaptation in Natural and Artificial Systems},
  author={Holland, John H.},
  year={1975},
  publisher={University of Michigan Press},
  address={Ann Arbor, MI, USA},
}

@book{goldberg1989genetic,
  title={Genetic Algorithms in Search, Optimization and Machine Learning},
  author={Goldberg, David E.},
  year={1989},
  publisher={Addison-Wesley},
  address={Reading, MA, USA},
  isbn={0-201-15767-5}
}

@book{schwefel1995evolution,
  title={Evolution and Optimum Seeking},
  author={Schwefel, Hans-Paul},
  year={1995},
  publisher={Wiley},
  address={Chichester, UK},
  isbn={0-471-96004-4}
}

@article{back1996evolutionary,
  title={Evolutionary Algorithms in Theory and Practice},
  author={B{\"a}ck, Thomas},
  journal={Oxford University Press},
  year={1996},
  volume={1},
  number={1},
  pages={361},
  doi={10.1093/acprof:oso/9780195099717.001.0001}
}

@Inbook{Dasgupta1997,
author="Dasgupta, Dipankar
and Michalewicz, Zbigniew",
editor="Dasgupta, Dipankar
and Michalewicz, Zbigniew",
title="Evolutionary Algorithms --- An Overview",
bookTitle="Evolutionary Algorithms in Engineering Applications",
year="1997",
publisher="Springer Berlin Heidelberg",
address="Berlin, Heidelberg",
pages="3--28",
isbn="978-3-662-03423-1",
doi="10.1007/978-3-662-03423-1_1",
url="https://doi.org/10.1007/978-3-662-03423-1_1"
}

@misc{hansen2019pycma,
  author       = {Nikolaus Hansen and Youhei Akimoto and Petr Baudis},
  title        = {{CMA-ES/pycma} on {G}ithub},
  howpublished = {Zenodo, DOI:10.5281/zenodo.2559634},
  month        = feb,
  year         = 2019,
  doi          = {10.5281/zenodo.2559634},
  url          = {https://doi.org/10.5281/zenodo.2559634},
}

@article{Leviant2022quantumcapacity,
  doi = {10.22331/q-2022-09-29-821},
  url = {https://doi.org/10.22331/q-2022-09-29-821},
  title = {Quantum capacity and codes for the bosonic loss-dephasing channel},
  author = {Leviant, Peter and Xu, Qian and Jiang, Liang and Rosenblum, Serge},
  journal = {{Quantum}},
  issn = {2521-327X},
  publisher = {{Verein zur F{\"{o}}rderung des Open Access Publizierens in den Quantenwissenschaften}},
  volume = {6},
  pages = {821},
  month = sep,
  year = {2022}
}

@article{Mele2024quantumcom_lossdephasing,
  title = {Quantum communication on the bosonic loss-dephasing channel},
  author = {Mele, Francesco Anna and Salek, Farzin and Giovannetti, Vittorio and Lami, Ludovico},
  journal = {Phys. Rev. A},
  volume = {110},
  issue = {1},
  pages = {012460},
  numpages = {26},
  year = {2024},
  month = {Jul},
  publisher = {American Physical Society},
  doi = {10.1103/PhysRevA.110.012460},
  url = {https://link.aps.org/doi/10.1103/PhysRevA.110.012460}
}

@article{Cochrane1999cat1,
  title = {Macroscopically distinct quantum-superposition states as a bosonic code for amplitude damping},
  author = {Cochrane, P. T. and Milburn, G. J. and Munro, W. J.},
  journal = {Phys. Rev. A},
  volume = {59},
  issue = {4},
  pages = {2631--2634},
  numpages = {0},
  year = {1999},
  month = {Apr},
  publisher = {American Physical Society},
  doi = {10.1103/PhysRevA.59.2631},
  url = {https://link.aps.org/doi/10.1103/PhysRevA.59.2631}
}

@article{Leghtas2013cat2,
  title = {Hardware-Efficient Autonomous Quantum Memory Protection},
  author = {Leghtas, Zaki and Kirchmair, Gerhard and Vlastakis, Brian and Schoelkopf, Robert J. and Devoret, Michel H. and Mirrahimi, Mazyar},
  journal = {Phys. Rev. Lett.},
  volume = {111},
  issue = {12},
  pages = {120501},
  numpages = {5},
  year = {2013},
  month = {Sep},
  publisher = {American Physical Society},
  doi = {10.1103/PhysRevLett.111.120501},
  url = {https://link.aps.org/doi/10.1103/PhysRevLett.111.120501}
}

@article{Michael2016binomial,
  title = {New Class of Quantum Error-Correcting Codes for a Bosonic Mode},
  author = {Michael, Marios H. and Silveri, Matti and Brierley, R. T. and Albert, Victor V. and Salmilehto, Juha and Jiang, Liang and Girvin, S. M.},
  journal = {Phys. Rev. X},
  volume = {6},
  issue = {3},
  pages = {031006},
  numpages = {26},
  year = {2016},
  month = {Jul},
  publisher = {American Physical Society},
  doi = {10.1103/PhysRevX.6.031006},
  url = {https://link.aps.org/doi/10.1103/PhysRevX.6.031006}
}

@Article{Ofek2016catexp,
author={Ofek, Nissim
and Petrenko, Andrei
and Heeres, Reinier
and Reinhold, Philip
and Leghtas, Zaki
and Vlastakis, Brian
and Liu, Yehan
and Frunzio, Luigi
and Girvin, S. M.
and Jiang, L.
and Mirrahimi, Mazyar
and Devoret, M. H.
and Schoelkopf, R. J.},
title={Extending the lifetime of a quantum bit with error correction in superconducting circuits},
journal={Nature},
year={2016},
month={Aug},
day={01},
volume={536},
number={7617},
pages={441-445},
issn={1476-4687},
doi={10.1038/nature18949},
url={https://doi.org/10.1038/nature18949}
}

@Article{Ni2023binomialexp,
author={Ni, Zhongchu
and Li, Sai
and Deng, Xiaowei
and Cai, Yanyan
and Zhang, Libo
and Wang, Weiting
and Yang, Zhen-Biao
and Yu, Haifeng
and Yan, Fei
and Liu, Song
and Zou, Chang-Ling
and Sun, Luyan
and Zheng, Shi-Biao
and Xu, Yuan
and Yu, Dapeng},
title={Beating the break-even point with a discrete-variable-encoded logical qubit},
journal={Nature},
year={2023},
month={Apr},
day={01},
volume={616},
number={7955},
pages={56-60},
issn={1476-4687},
doi={10.1038/s41586-023-05784-4},
url={https://doi.org/10.1038/s41586-023-05784-4}
}

@Article{Sivak2023,
author={Sivak, V. V.
and Eickbusch, A.
and Royer, B.
and Singh, S.
and Tsioutsios, I.
and Ganjam, S.
and Miano, A.
and Brock, B. L.
and Ding, A. Z.
and Frunzio, L.
and Girvin, S. M.
and Schoelkopf, R. J.
and Devoret, M. H.},
title={Real-time quantum error correction beyond break-even},
journal={Nature},
year={2023},
month={Apr},
day={01},
volume={616},
number={7955},
pages={50-55},
issn={1476-4687},
doi={10.1038/s41586-023-05782-6},
url={https://doi.org/10.1038/s41586-023-05782-6}
}

@Article{Brock2025,
author={Brock, Benjamin L.
and Singh, Shraddha
and Eickbusch, Alec
and Sivak, Volodymyr V.
and Ding, Andy Z.
and Frunzio, Luigi
and Girvin, Steven M.
and Devoret, Michel H.},
title={Quantum error correction of qudits beyond break-even},
journal={Nature},
year={2025},
month={May},
day={01},
volume={641},
number={8063},
pages={612-618},
issn={1476-4687},
doi={10.1038/s41586-025-08899-y},
url={https://doi.org/10.1038/s41586-025-08899-y}
}

@Article{Provazník2022Taming,
author={Provazn{\'i}k, Jan
and Filip, Radim
and Marek, Petr},
title={Taming numerical errors in simulations of continuous variable non-Gaussian state preparation},
journal={Sci. Rep.},
year={2022},
month={Oct},
day={04},
volume={12},
number={1},
pages={16574},
issn={2045-2322},
doi={10.1038/s41598-022-19506-9},
url={https://doi.org/10.1038/s41598-022-19506-9}
}

@ARTICLE{Ouyang2021NPtradeoff,
  author={Ouyang, Yingkai and Campbell, Earl T.},
  journal={IEEE Trans. Inf. Theory}, 
  title={Trade-Offs on Number and Phase Shift Resilience in Bosonic Quantum Codes}, 
  year={2021},
  volume={67},
  number={10},
  pages={6644-6652},
  doi={10.1109/TIT.2021.3102873}}

@article{
Teoh2023Dualencoding,
author = {James D. Teoh  and Patrick Winkel  and Harshvardhan K. Babla  and Benjamin J. Chapman  and Jahan Claes  and Stijn J. de Graaf  and John W. O. Garmon  and William D. Kalfus  and Yao Lu  and Aniket Maiti  and Kaavya Sahay  and Neel Thakur  and Takahiro Tsunoda  and Sophia H. Xue  and Luigi Frunzio  and Steven M. Girvin  and Shruti Puri  and Robert J. Schoelkopf },
title = {Dual-rail encoding with superconducting cavities},
journal = {Proc. Natl. Acad. Sci. U.S.A.},
volume = {120},
number = {41},
pages = {e2221736120},
year = {2023},
doi = {10.1073/pnas.2221736120},
URL = {https://www.pnas.org/doi/abs/10.1073/pnas.2221736120},
}

@Article{Cai2024,
author={Cai, Weizhou
and Mu, Xianghao
and Wang, Weiting
and Zhou, Jie
and Ma, Yuwei
and Pan, Xiaoxuan
and Hua, Ziyue
and Liu, Xinyu
and Xue, Guangming
and Yu, Haifeng
and Wang, Haiyan
and Song, Yipu
and Zou, Chang-Ling
and Sun, Luyan},
title={Protecting entanglement between logical qubits via quantum error correction},
journal={Nat. Phys.},
year={2024},
month={Jun},
day={01},
volume={20},
number={6},
pages={1022-1026},
issn={1745-2481},
doi={10.1038/s41567-024-02446-8},
url={https://doi.org/10.1038/s41567-024-02446-8}
}

@article{Lloyd1998Analog,
  title = {Analog Quantum Error Correction},
  author = {Lloyd, Seth and Slotine, Jean-Jacques E.},
  journal = {Phys. Rev. Lett.},
  volume = {80},
  issue = {18},
  pages = {4088--4091},
  numpages = {0},
  year = {1998},
  month = {May},
  publisher = {American Physical Society},
  doi = {10.1103/PhysRevLett.80.4088},
  url = {https://link.aps.org/doi/10.1103/PhysRevLett.80.4088}
}

@article{Braunstein1998QECforQV,
  title = {Error Correction for Continuous Quantum Variables},
  author = {Braunstein, Samuel L.},
  journal = {Phys. Rev. Lett.},
  volume = {80},
  issue = {18},
  pages = {4084--4087},
  numpages = {0},
  year = {1998},
  month = {May},
  publisher = {American Physical Society},
  doi = {10.1103/PhysRevLett.80.4084},
  url = {https://link.aps.org/doi/10.1103/PhysRevLett.80.4084}
}

@article{Hayden_2016,
doi = {10.1088/1367-2630/18/8/083043},
url = {https://doi.org/10.1088/1367-2630/18/8/083043},
year = {2016},
month = {aug},
publisher = {IOP Publishing},
volume = {18},
number = {8},
pages = {083043},
author = {Hayden, Patrick and Nezami, Sepehr and Salton, Grant and Sanders, Barry C},
title = {Spacetime replication of continuous variable quantum information},
journal = {New J. Phys.}
}

@article{Niset2008Feasible,
  title = {Experimentally Feasible Quantum Erasure-Correcting Code for Continuous Variables},
  author = {Niset, J. and Andersen, U. L. and Cerf, N. J.},
  journal = {Phys. Rev. Lett.},
  volume = {101},
  issue = {13},
  pages = {130503},
  numpages = {4},
  year = {2008},
  month = {Sep},
  publisher = {American Physical Society},
  doi = {10.1103/PhysRevLett.101.130503},
  url = {https://link.aps.org/doi/10.1103/PhysRevLett.101.130503}
}

@article{Chuang1997Bosoniccodes,
  title = {Bosonic quantum codes for amplitude damping},
  author = {Chuang, Isaac L. and Leung, Debbie W. and Yamamoto, Yoshihisa},
  journal = {Phys. Rev. A},
  volume = {56},
  issue = {2},
  pages = {1114--1125},
  numpages = {0},
  year = {1997},
  month = {Aug},
  publisher = {American Physical Society},
  doi = {10.1103/PhysRevA.56.1114},
  url = {https://link.aps.org/doi/10.1103/PhysRevA.56.1114}
}

@article{Bergmann2016NOONstates,
  title = {Quantum error correction against photon loss using NOON states},
  author = {Bergmann, Marcel and van Loock, Peter},
  journal = {Phys. Rev. A},
  volume = {94},
  issue = {1},
  pages = {012311},
  numpages = {13},
  year = {2016},
  month = {Jul},
  publisher = {American Physical Society},
  doi = {10.1103/PhysRevA.94.012311},
  url = {https://link.aps.org/doi/10.1103/PhysRevA.94.012311}
}

@article{Niu2018sym_ope,
  title = {Hardware-efficient bosonic quantum error-correcting codes based on symmetry operators},
  author = {Niu, Murphy Yuezhen and Chuang, Isaac L. and Shapiro, Jeffrey H.},
  journal = {Phys. Rev. A},
  volume = {97},
  issue = {3},
  pages = {032323},
  numpages = {21},
  year = {2018},
  month = {Mar},
  publisher = {American Physical Society},
  doi = {10.1103/PhysRevA.97.032323},
  url = {https://link.aps.org/doi/10.1103/PhysRevA.97.032323}
}

@article{BRADY2024100496,
title = {Advances in bosonic quantum error correction with Gottesman–Kitaev–Preskill Codes: Theory, engineering and applications},
journal = {Prog. Quantum Electron.},
volume = {93},
pages = {100496},
year = {2024},
issn = {0079-6727},
doi = {https://doi.org/10.1016/j.pquantelec.2023.100496},
url = {https://www.sciencedirect.com/science/article/pii/S0079672723000459},
author = {Anthony J. Brady and Alec Eickbusch and Shraddha Singh and Jing Wu and Quntao Zhuang}
}

@article{Bashmakova2025squeezedFock,
  title = {Bosonic quantum error correction using squeezed Fock states},
  author = {Bashmakova, E. N. and Korolev, S. B. and Golubeva, T. Yu.},
  journal = {Phys. Rev. A},
  volume = {112},
  issue = {3},
  pages = {032434},
  numpages = {13},
  year = {2025},
  month = {Sep},
  publisher = {American Physical Society},
  doi = {10.1103/97yt-nzg2},
  url = {https://link.aps.org/doi/10.1103/97yt-nzg2}
}

@article{Zeng2023ApproxRL,
  title = {Approximate Autonomous Quantum Error Correction with Reinforcement Learning},
  author = {Zeng, Yexiong and Zhou, Zheng-Yang and Rinaldi, Enrico and Gneiting, Clemens and Nori, Franco},
  journal = {Phys. Rev. Lett.},
  volume = {131},
  issue = {5},
  pages = {050601},
  numpages = {8},
  year = {2023},
  month = {Jul},
  publisher = {American Physical Society},
  doi = {10.1103/PhysRevLett.131.050601},
  url = {https://link.aps.org/doi/10.1103/PhysRevLett.131.050601}
}

@article{Zeng2025ApproxGKP,
  title = {Neural-Network-Based Design of Approximate Gottesman-Kitaev-Preskill Code},
  author = {Zeng, Yexiong and Qin, Wei and Chen, Ye-Hong and Gneiting, Clemens and Nori, Franco},
  journal = {Phys. Rev. Lett.},
  volume = {134},
  issue = {6},
  pages = {060601},
  numpages = {7},
  year = {2025},
  month = {Feb},
  publisher = {American Physical Society},
  doi = {10.1103/PhysRevLett.134.060601},
  url = {https://link.aps.org/doi/10.1103/PhysRevLett.134.060601}
}

@Article{Jain2024sphericalcodes,
author={Jain, Shubham P.
and Iosue, Joseph T.
and Barg, Alexander
and Albert, Victor V.},
title={Quantum spherical codes},
journal={Nat. Phys.},
year={2024},
month={Aug},
day={01},
volume={20},
number={8},
pages={1300-1305},
issn={1745-2481},
doi={10.1038/s41567-024-02496-y},
url={https://doi.org/10.1038/s41567-024-02496-y}
}

@article{choi1975completely,
  title={Completely positive linear maps on complex matrices},
  author={Choi, Man-Duen},
  journal={Linear Algebra and its Applications},
  volume={10},
  number={3},
  pages={285--290},
  year={1975},
  publisher={Elsevier},
  doi={10.1016/0024-3795(75)90075-0}
}

@inproceedings{kitaev1997quantum,
  title={Quantum computations: algorithms and error correction},
  author={Kitaev, Alexei Y.},
  booktitle={Russian Mathematical Surveys},
  volume={52},
  number={6},
  pages={1191--1249},
  year={1997},
  url = {https://doi.org/10.1070/RM1997v052n06ABEH002155},
  doi = {10.1070/RM1997v052n06ABEH002155},
}

@article{schumacher1996entanglement,
  title={Sending entanglement through noisy quantum channels},
  author={Schumacher, Benjamin},
  journal={Phys. Rev. A},
  volume={54},
  number={4},
  pages={2614--2628},
  year={1996},
  publisher={American Physical Society},
  doi={10.1103/PhysRevA.54.2614}
}

@article{Fletcher2007OptRecovery,
  title = {Optimum quantum error recovery using semidefinite programming},
  author = {Fletcher, Andrew S. and Shor, Peter W. and Win, Moe Z.},
  journal = {Phys. Rev. A},
  volume = {75},
  issue = {1},
  pages = {012338},
  numpages = {7},
  year = {2007},
  month = {Jan},
  publisher = {American Physical Society},
  doi = {10.1103/PhysRevA.75.012338},
  url = {https://link.aps.org/doi/10.1103/PhysRevA.75.012338}
}

@article{Reimpell2005IterativeOpt,
  title = {Iterative Optimization of Quantum Error Correcting Codes},
  author = {Reimpell, M. and Werner, R. F.},
  journal = {Phys. Rev. Lett.},
  volume = {94},
  issue = {8},
  pages = {080501},
  numpages = {4},
  year = {2005},
  month = {Mar},
  publisher = {American Physical Society},
  doi = {10.1103/PhysRevLett.94.080501},
  url = {https://link.aps.org/doi/10.1103/PhysRevLett.94.080501}
}

@article{watrous2009semidefinite,
  title={Semidefinite programs for completely bounded norms},
  author={Watrous, John},
  journal={Theory of Computing},
  volume={5},
  number={11},
  pages={217--238},
  year={2009},
  doi={10.4086/toc.2009.v005a011}
}

@article{kosut2008robust,
  title={Robust quantum error correction via convex optimization},
  author={Kosut, Robert L. and Shabani, Alireza and Lidar, Daniel A.},
  journal={Phys. Rev. Lett.},
  volume={100},
  number={2},
  pages={020502},
  year={2008},
  publisher={American Physical Society},
  doi={10.1103/PhysRevLett.100.020502}
}

@article{Audenaert2002OptSDP1,
  title = {Optimizing completely positive maps using semidefinite programming},
  author = {Audenaert, Koenraad and De Moor, Bart},
  journal = {Phys. Rev. A},
  volume = {65},
  issue = {3},
  pages = {030302},
  numpages = {4},
  year = {2002},
  month = {Feb},
  publisher = {American Physical Society},
  doi = {10.1103/PhysRevA.65.030302},
  url = {https://link.aps.org/doi/10.1103/PhysRevA.65.030302}
}

@article{Mironowicz2024SDPandQI,
doi = {10.1088/1751-8121/ad2b85},
url = {https://doi.org/10.1088/1751-8121/ad2b85},
year = {2024},
month = {apr},
publisher = {IOP Publishing},
volume = {57},
number = {16},
pages = {163002},
author = {Mironowicz, Piotr},
title = {Semi-definite programming and quantum information},
journal = {J. Phys. A: Math. Theor.}
}

@article{gh1c-xyn1,
  title = {Performance and Achievable Rates of the Gottesman-Kitaev-Preskill Code for Pure-Loss and Amplification Channels},
  author = {Zheng, Guo and He, Wenhao and Lee, Gideon and Noh, Kyungjoo and Jiang, Liang},
  journal = {PRX Quantum},
  volume = {6},
  issue = {3},
  pages = {030314},
  numpages = {31},
  year = {2025},
  month = {Jul},
  publisher = {American Physical Society},
  doi = {10.1103/gh1c-xyn1},
  url = {https://link.aps.org/doi/10.1103/gh1c-xyn1}
}

\end{document}